\documentclass[UKenglish,cleveref,numberwithinsect,thm-restate,a4paper,11pt]{article}
\usepackage[utf8]{inputenc}
\usepackage{a4wide}
\usepackage{hyperref}
\hypersetup{colorlinks=true,citecolor=blue,linkcolor=blue,urlcolor=blue}
\usepackage{amsmath}
\usepackage{amssymb}
\usepackage{amsthm}
\usepackage{enumerate}
\usepackage{comment}
\usepackage{thmtools, thm-restate}
\usepackage{graphicx}
\usepackage{blkarray}
\usepackage{tikz, tikz-cd}
\usetikzlibrary{matrix}
\usetikzlibrary{shapes}
\usetikzlibrary{arrows,decorations.markings, tikzmark}
\usetikzlibrary{decorations.pathreplacing,calligraphy,backgrounds}
\usetikzlibrary{arrows.meta,arrows}
\usepackage{xcolor}
\usepackage{mathtools}
\usepackage{stmaryrd}
\usepackage{tabularx}
\usepackage{xspace}
\usepackage{booktabs}
\usepackage{multirow}

\theoremstyle{plain}
\newtheorem{theorem}{Theorem}
\newtheorem*{theorem*}{Theorem}
\newtheorem{corollary}[theorem]{Corollary}
\newtheorem{lemma}[theorem]{Lemma}
\newtheorem{observation}[theorem]{Observation}
\newtheorem{definition}[theorem]{Definition}

\newtheorem{remark}[theorem]{Remark}

\newcommand{\subsprob}{\textsc{Sub}}
\newcommand{\indsubsprob}{\textsc{IndSub}}
\newcommand{\homsprob}{\textsc{Hom}}
\newcommand{\homs}[2]{\mathsf{Hom}(#1 \to #2)}
\newcommand{\surhoms}[2]{\mathsf{SurHom}(#1 \to #2)}

\newcommand{\subs}[2]{\mathsf{Sub}(#1 \to #2)}
\newcommand{\indsubs}[2]{\mathsf{IndSub}(#1 \to #2)}

\newcommand{\calG}{\mathcal{G}}

\newcommand{\tw}{\mathsf{tw}}
\newcommand{\id}{\mathsf{id}}

\newcommand{\W}{\mathrm{W}}

\newcommand{\N}{\mathbb{N}}
\newcommand{\Q}{\mathbb{Q}}
\newcommand{\No}{\N_0}

\newcommand{\scH}{\mathcal{H}}
\newcommand{\scU}{\mathcal{U}}
\newcommand{\scG}{\mathcal{G}}

\newcommand{\scI}{\mathcal{I}}

\newcommand{\ccStyle}[1]{\mathrm{#1}}
\newcommand{\ccP}{\ccStyle{P}}
\newcommand{\ccSharpP}{\#\ccStyle{P}}
\newcommand{\ccFPT}{\ccStyle{FPT}}
\newcommand{\ccW}[1]{\ccStyle{W}\text{[#1]}}
\newcommand{\ccSharpW}[1]{\#\ccStyle{W}\text{[#1]}}

\newcommand{\coledgesubs}[2]{\mbox{\ensuremath{\mathsf{ColEdgeSub}(#1 \to #2)}}}

\newcommand{\clique}{\textsc{Clique}}
\newcommand{\is}{\textsc{IndSet}}
\newcommand{\colis}{\textsc{ColIndSet}}
\newcommand{\match}{\textsc{Match}}
\newcommand{\colmatch}{\textsc{ColMatch}}
\newcommand{\cphomsprob}{\text{\sc{cp-Hom}}}
\newcommand{\bind}{\beta_{\mathsf{ind}}}
\newcommand{\m}{{\mathrm{m}}}
\newcommand{\mind}{\m_{\mathsf{ind}}}
\newcommand{\fptred}{\leq^{\mathsf{FPT}}}
\newcommand{\all}{{\mathcal{U}}}
\def\fracture#1#2{\ensuremath{#1\raisebox{.2ex}{\rotatebox[origin=c]{-15}{$\sharp$}}#2}}

\newcommand{\redc}[2][red,fill=red]{\tikz[baseline=-0.5ex]\draw[#1,radius=#2] (0,0) circle ;}
\newcommand{\greenc}[2][green!80!blue,fill=green!80!blue]{\tikz[baseline=-0.5ex]\draw[#1,radius=#2] (0,0) circle ;}
\newcommand{\bluec}[2][blue,fill=blue]{\tikz[baseline=-0.5ex]\draw[#1,radius=#2] (0,0) circle ;}
\newcommand{\yellowc}[2][yellow!50!orange,fill=yellow!50!orange]{\tikz[baseline=-0.5ex]\draw[#1,radius=#2] (0,0) circle ;}
\newcommand{\blackc}[2][black,fill=black]{\tikz[baseline=-0.5ex]\draw[#1,radius=#2] (0,0) circle ;}
\newcommand{\brownc}[2][cyan,fill=cyan]{\tikz[baseline=-0.5ex]\draw[#1,radius=#2] (0,0) circle ;}
\usetikzlibrary{conference}
\usetikzlibrary{cd}

\title{\vspace{-10mm} Counting Subgraphs in Somewhere Dense Graphs\footnote{This research was funded in whole, or in part, by EPSRC grant EP/V032305/1. For the purpose of Open Access, the authors have applied a CC BY public copyright licence to any Author Accepted Manuscript version arising from this submission. All data is provided in full in the results section of this paper.}}

\author{Marco Bressan \\ Department of Computer Science\\ University of Milan\\ Italy 
\and
Leslie Ann Goldberg \\ Department of Computer Science \\ University of Oxford\\ United Kingdom 
\and
Kitty Meeks \\ School of Computing Science \\ University of Glasgow \\ United Kingdom
\and
Marc Roth \\ School of Electronic Engineering and Computer Science \\ Queen Mary University of London\\ United Kingdom 
}
\date{}

\begin{document}

\maketitle

\begin{abstract}
    We study the problems of counting copies and induced copies of a small pattern graph $H$ in a large host graph $G$. Recent work fully classified the complexity of those problems according to structural restrictions on the patterns $H$. In this work, we address the more challenging task of analysing the complexity for restricted patterns \emph{and} restricted hosts. Specifically we ask which families of allowed patterns and hosts imply fixed-parameter tractability, i.e., the existence of an algorithm running in time $f(H)\cdot |G|^{O(1)}$ for some computable function $f$.
    Our main results present exhaustive and explicit complexity classifications for families that satisfy natural closure properties. Among others, we identify the problems of counting small matchings and independent sets in subgraph-closed graph classes $\mathcal{G}$ as our central objects of study and establish the following crisp dichotomies as consequences of the Exponential Time Hypothesis:
    \begin{itemize}
        \item Counting $k$-matchings in a graph $G\in\mathcal{G}$ is fixed-parameter tractable if and only if $\mathcal{G}$ is nowhere dense.
        \item Counting $k$-independent sets in a graph $G\in\mathcal{G}$ is fixed-parameter tractable if and only if $\mathcal{G}$ is nowhere dense.
    \end{itemize}
    Moreover, we obtain almost tight conditional lower bounds if $\mathcal{G}$ is somewhere dense, i.e., not nowhere dense. These base cases of our classifications subsume a wide variety of previous results on the matching and independent set problem, such as counting $k$-matchings in bipartite graphs (Curticapean, Marx; FOCS 14), in $F$-colourable graphs (Roth, Wellnitz; SODA 20), and in degenerate graphs (Bressan, Roth; FOCS 21), as well as counting $k$-independent sets in bipartite graphs (Curticapean et al.; Algorithmica 19).
    
    At the same time our proofs are much simpler: using structural characterisations of somewhere dense graphs, we show that a colourful version of a recent breakthrough technique for analysing pattern counting problems (Curticapean, Dell, Marx; STOC 17) applies to \emph{any} subgraph-closed somewhere dense class of graphs, yielding a unified view of our current understanding of the complexity of subgraph counting.
\end{abstract}

\section{Introduction}
We study the following subgraph counting problem: given two graphs $H$ and $G$, compute the number of copies of $H$ in $G$. For several decades this problem has received widespread attention from the theoretical community, leading to a rich algorithmic toolbox that draws from different techniques~\cite{NesetrilP85,AlonYZ95,BubleyD97,JerrumSV04} and to deep structural results in parameterised complexity theory~\cite{FlumG04,CurticapeanDM17}. Since it was discovered that subgraph counts reveal global properties of complex networks~\cite{Miloetal02,Miloetal04}, subgraph counting has also found several applications in fields such as biology~\cite{Nogaetat08,Schilleretal15} genetics~\cite{Tranetal13}, phylogeny~\cite{Kuchaievetal10}, and data mining~\cite{Babis17}.
Unfortunately, the subgraph counting problem is in general intractable, since it contains as special cases hard problems such as \textsc{Clique}. This does not mean however that the problem is \emph{always} intractable; it just means that it is tractable when the pattern $H$ is restricted to certain graph families. Identifying these families of patterns that are efficiently countable has been a key question for the last twenty years. A long stream of research eventually showed that, unless standard conjectures fail, subgraph counting is tractable only for very restricted families of patterns~\cite{FlumG04,DalmauJ04,ChenTW08,CurticapeanM14,JerrumM15,Meeks16,CurticapeanDM17,RothSW20,FockeR22}.

To circumvent this ``wall of intractability'', in this work we restrict both the family of the pattern $H$ and the family of the host $G$. Formally, given two classes of graphs $\mathcal{H}$ and $\mathcal{G}$, we study the problems $\#\subsprob(\mathcal{H}\to\mathcal{G})$,  $\#\indsubsprob(\mathcal{H}\to\mathcal{G})$, and $\#\homsprob(\mathcal{H}\to\mathcal{G})$, defined as follows. For all of them, the input is a pair $(H,G)$ with $H\in\mathcal{H}$ and $G\in\mathcal{G}$. The outputs are respectively the number of subgraphs of $G$ isomorphic to $H$, denoted by $\#\subs{H}{G}$, the number of induced subgraphs of $G$ isomorphic to $H$, denoted by $\#\indsubs{H}{G}$, and the number of homomorphisms (edge-preserving maps) from $H$ to $G$, denoted by $\#\homs{H}{G}$.
Our goal is to determine for which $\scH$ and $\scG$ these three problems are tractable. To formalize what we mean by tractable, we adopt the framework of parameterized complexity~\cite{CyganFKLMPPS15}: we say that a problem is \emph{fixed-parameter tractable}, or in the class $\ccFPT$, if it is solvable in time $f(|H|)\cdot |G|^{O(1)}$ for some computable function $f$ (see Section~\ref{sec:prelims} for a complete introduction). For instance, we consider as tractable a running time of $2^{O(|H|)} \cdot |G|$ but not one of $|G|^{O(|H|)}$. This captures the intuition that $H$ is ``small'' compared to $G$, and is the main theoretical framework for subgraph counting~\cite{FlumG04}. Thus, the goal of this work is understanding the fixed-parameter tractability of $\#\subsprob(\mathcal{H}\to\mathcal{G})$,  $\#\indsubsprob(\mathcal{H}\to\mathcal{G})$, and $\#\homsprob(\mathcal{H}\to\mathcal{G})$ as a function of $\scH$ and $\scG$. Moreover, when those problems are not fixed-parameter tractable we aim to show that they are hard for the complexity class $\#\W[1]$, which can be thought of as the equivalent of $\mathrm{NP}$ for parameterized counting. 

We first briefly discuss which properties of $\scG$ are worthy of attention. When $\mathcal{G}$ is the class of all graphs, it is well known that each of the three problems  is either $\ccFPT$ or $\ccSharpW{1}$-hard depending on whether certain structural parameters of $\scH$ (such as treewidth or vertex cover number) are bounded.
Thus, when $\mathcal{G}$ is the class of all graphs, the problem is solved.
However, when $\scG$ is arbitrary, no such characterization is known. This is partly due to the fact that ``natural'' structural properties related to subgraph counting are harder to find for $\scG$ than for $\scH$; subgraph counting algorithms themselves usually exploit the structure of $H$ but not that of $G$ (think of tree decompositions).
There is however one deep structural property that, if held by $\scG$, yields tractability: the property of being \emph{nowhere dense}, introduced by Ne{\v{s}}et{\v{r}}il and Ossona de Mendez~\cite{NesetrildM11}. In a nutshell $\scG$ is nowhere dense if, for all $r \in \No$, its members do not contain as subgraphs the $r$-subdivisions of arbitrarily large cliques; it can be shown that this generalizes several natural definitions of sparsity, including having bounded degree or bounded local treewidth, or excluding some topological minor.
In a remarkable result, Ne{\v{s}}et{\v{r}}il and Ossona de Mendez proved:\footnote{In the realm of decision problems, an even more general meta-theorem is known for first-order model-checking on nowhere dense graphs~\cite{GroheKS17}.}

\vbox{
\begin{theorem}[Theorem 18.9 in~\cite{Sparsity}]\label{thm:sparsity_intro}
If $\mathcal{G}$ is nowhere dense then $\#\homsprob(\mathcal{H} \to \mathcal{G})$, $\#\subsprob(\mathcal{H} \to \mathcal{G})$, and $\#\indsubsprob(\mathcal{H} \to \mathcal{G})$ are fixed-parameter tractable and can be solved in time
$f(|H|)\cdot |V(G)|^{1+o(1)}$ for some computable function $f$.
\end{theorem}}

\noindent Thus the case of nowhere dense $\scG$ is closed, and we can focus on its complement --- the case where $\scG$ is \emph{somewhere dense}. Hence the question studied in this work is: when are $\#\subsprob(\mathcal{H}\to\mathcal{G})$,  $\#\indsubsprob(\mathcal{H}\to\mathcal{G})$, and $\#\homsprob(\mathcal{H}\to\mathcal{G})$ fixed-parameter tractable, provided $\scG$ is somewhere dense?

\subsection{Our Results}\label{sec:results}
We prove dichotomies for $\#\subsprob(\mathcal{H}\to\mathcal{G})$,  $\#\indsubsprob(\mathcal{H}\to\mathcal{G})$, and $\#\homsprob(\mathcal{H}\to\mathcal{G})$ into $\ccFPT$ and $\ccSharpW{1}$-hard cases, assuming that $\scG$ is somewhere dense. 
It is known~\cite{RothW20} that a fully general dichotomy is  impossible even assuming that $\scG$ is somewhere dense; thus we focus on the natural cases where $\scH$ and/or $\scG$ are monotone (closed under taking subgraphs) or hereditary (closed under taking induced subgraphs).
Our dichotomies are expressed in terms of the finiteness of combinatorial parameters of $\scH$ and $\scG$, such as their clique number or their induced matching number.  
Existing complexity dichotomies for subgraph counting  are based on using interpolation to evaluate linear combinations of homomorphism counts~\cite{CurticapeanDM17}. This technique has been exploited for families of host graphs that are closed under tensoring --- the closure is used to create new instances for the interpolation. 
The host graphs in our dichotomy theorems do not have  this closure property. Nevertheless, we obtain a dichotomy for all somewhere dense classes using a combination of techniques involving graph fractures and colourings.

The rest of this section presents our main conceptual contribution (Section~\ref{sec:concept}), gives a detailed walk-through of our complexity dichotomies (Section~\ref{sub:subs}, Section~\ref{sub:inds}, Section~\ref{sub:homs}), provides some context (Section~\ref{sec:context}), and overviews the techniques behind our proofs (Section~\ref{sec:tech}). For full proofs of our claims see Section~\ref{sec:prelims} onward.

\paragraph{Basic preliminaries.} We concisely state some necessary definitions and observations which are given in more detail in Section~\ref{sec:prelims}. We denote by $\all$ the class of all graphs. We denote by $\omega(G),\alpha(G),\beta(G)$, and $\m(G)$ respectively the clique, independence, biclique, and matching number of a graph $G$. The notation extends to graph classes by taking the supremum over their elements. Induced versions of those quantities are identified by the subscript $_{\mathsf{ind}}$ (for instance, $\mind$ denotes the induced matching number). $G^r$ denotes the $r$-subdivision of $G$, and $F \times G$ denotes the tensor product of $F$ and $G$.
All of our lower bounds assume the Exponential Time Hypothesis (ETH)~\cite{ImpagliazzoP01}; and most of them rule out algorithms running in time $f(k)\cdot n^{o(k/\log k)}$ for any function $f$, and are therefore tight except possibly for a $O(\log k)$ factor in the exponent.\footnote{This $O(\log k)$ gap is not an artifact of our proofs, but a consequence of the well-known open problem ``Can you beat treewidth?''~\cite{Marx10,Marx13}.} 
All of our $\#\W[1]$-hardness results are actually $\#\W[1]$-completeness results; this holds because $\#\subsprob(\mathcal{H}\to\mathcal{G})$,  $\#\indsubsprob(\mathcal{H}\to\mathcal{G})$, and $\#\homsprob(\mathcal{H}\to\mathcal{G})$ are always in $\#\W[1]$ due to a characterisation of $\#\W[1]$ via parameterised model-counting problems (see~\cite[Chapter 14]{FlumG06}).

\subsubsection{Simpler Hardness Proofs for More Graph Families}\label{sec:concept}
Our first and most conceptual contribution is a novel approach to proving hardness of parameterized subgraph counting problems for somewhere-dense families of host graphs. This approach allows us to significantly generalize existing results while simultaneously yielding surprisingly simpler proofs.

The starting point is the observation that proving intractability results for parameterized counting problems is discouragingly difficult, as it often requires tedious and involved arguments. For instance, after Flum and Grohe conjectured that counting $k$-matchings is $\#\W[1]$-hard~\cite{FlumG04}, the first proof required nine years and relied on sophisticated algebraic techniques~\cite{Curticapean13}. This partially changed in 2017 when Curticapean, Dell and Marx~\cite{CurticapeanDM17}
showed how to express a subgraph count $\#\subs{H}{G}$ as linear combination of homomorphism counts $\sum_{F} a_F \cdot \#\homs{F}{G}$. They showed that computing this linear combination has the same complexity as computing the hardest term $\#\homs{F}{G}$ such that $a_F \ne 0$. A similar claim holds for induced subgraph counts as well. Thanks to this technique one can prove intractability of several subgraph counting problems, including for instance the problem of counting $k$-matchings.\footnote{In the field of database theory a similar technique expressing answers to unions of conjunctive queries as linear combinations of answers of conjunctive queries was independently discovered by Chen and Mengel~\cite{ChenM16}.}
These hardness results ultimately yielded complexity dichotomies for general subgraph counting problems, including notably $\#\subsprob(\mathcal{H}\to \mathcal G)$ and $\#\indsubsprob(\mathcal{H}\to \mathcal G)$ when $\mathcal G$ is the class of all graphs.

The technique of~\cite{CurticapeanDM17} does not work for proving hardness of $\#\subsprob(\mathcal{H}\to \mathcal G)$ and $\#\indsubsprob(\mathcal{H}\to \mathcal G)$ when $\mathcal G \neq \all$.
Indeed, one caveat of that technique is that the family of host graphs $\scG$ must satisfy certain conditions. One of those conditions is that $\scG$ is closed under tensoring, i.e., that $G \times G' \in \scG$ for all $G\in \mathcal{G}$ and all $G' \in \all$. The reason is that the interpolation relies on evaluating, say, $\subs{H}{G \times G_i}$ for several carefully chosen graphs $G_i$, with the goal of constructing a certain invertible system of linear equations; for this to yield a reduction towards counting patterns in graphs from $\mathcal G$, it is crucial that $G \times G_i \in \mathcal G$ for all such $G_i$ (Section~\ref{sec:tech} gives a concrete example using the problem of counting $k$-matchings). This is why the technique of~\cite{CurticapeanDM17} works smoothly for $\scG=\scU$; closure under tensoring holds trivially in that case.
But many other natural graph families $\mathcal G$ are not closed under tensoring, including somewhere dense ones (for instance, the family of $d$-degenerate graphs for any fixed integer $d\geq 2$). Until now, this has been the main obstacle towards proving hardness of subgraph counting for arbitrary somewhere dense graph families.
The central insight of our work is that this obstacle can be circumvented in a surprisingly simple way. Using well-established results from the theory of sparsity, we prove the following claim, which we explain in detail in Section~\ref{sec:tech}:
\begin{center}
\textit{Every monotone and somewhere dense class of graphs is closed \\under vertex-colourful tensor products of subdivided graphs.}
\end{center}
Ignoring for a moment its technicalities, this result allows us to lift the interpolation technique via graph tensors to \emph{any} monotone somewhere dense class of host graphs, including for instance the aforementioned class of $d$-degenerate graphs. In turn this yields complexity classifications for $\#\homsprob(\mathcal{H}\to\mathcal{G})$, $\#\subsprob(\mathcal{H}\to\mathcal{G})$, and $\#\indsubsprob(\mathcal{H}\to\mathcal{G})$ that subsume and significantly strengthen almost all classifications known in the literature (see below). Moreover, our approach yields simple and almost self-contained proofs, helping understand the underlying causes of the hardness.

\subsubsection{The Complexity of $\#\subsprob(\scH \to \scG)$}\label{sub:subs}
This section presents our results on the fixed-parameter tractability of $\#\subsprob(\scH \to \scG)$. We start by presenting a minimal\footnote{Minimal means that, for every class $\scH'$, $\#\subsprob(\scH' \to \scG)$ is intractable if and only if the monotone closure of $\scH'$ includes $\scH$. The same holds for $\#\indsubsprob$ with ``monotone'' replaced by ``hereditary''.} family $\scH$ for which 
hardness holds:
the family of all $k$-matchings (or $1$-regular graphs). In this case we also denote $\#\subsprob(\scH \to \scG)$ as $\#\match(\mathcal{G})$.
In the foundational work by Flum and Grohe~\cite{FlumG04}, $\#\match(\all)$ was identified as a central problem because of the significance of its classical counterpart (counting the number of perfect matchings); a series of works then identified $\#\match(\all)$ as the minimal intractable case~\cite{Curticapean13,CurticapeanM14,CurticapeanDM17}.
In this work, we show that $\#\match(\scG)$ is the minimal hard case for \emph{every} class $\scG$ that is monotone and somewhere dense:
\begin{theorem}\label{thm:intro_match}
Let $\mathcal{G}$ be a monotone class of graphs\footnote{We emphasize that we do not need our classes to be computable or recursively enumerable. This is due to the assumed closure properties of the classes.} and assume that ETH holds. Then $\#\match(\mathcal{G})$ is fixed-parameter tractable if and only if $\mathcal{G}$ is nowhere dense. More precisely, if $\mathcal{G}$ is nowhere dense then $\#\match(\mathcal{G})$ can be solved in time $f(k)\cdot|V(G)|^{1+o(1)}$ for some computable function~$f$; otherwise $\#\match(\mathcal{G})$ is $\#\W[1]$-hard and cannot be solved in time $f(k)\cdot |G|^{o(k/\log k)}$ for any function $f$. 
\end{theorem}
\noindent Theorem~\ref{thm:intro_match} subsumes the existing intractability results for counting $k$-matchings in bipartite graphs~\cite{CurticapeanM14}, in $F$-colourable graphs~\cite{RothW20}, in bipartite graphs with one-sided degree bounds~\cite{CurticapeanDR17}, and in degenerate graphs~\cite{BressanR21}. It also strengthens the latter result: while \cite{BressanR21} establishes hardness of counting $k$-matchings in $\ell$-degenerate graphs for $k+\ell$ as a parameter, Theorem~\ref{thm:intro_match} yields hardness for $d$-degenerate graphs for every fixed $d\geq 2$.\footnote{The class of all $d$-degenerate graphs is somewhere dense for all $d \ge 2$.}
Additionally, we show that Theorem~\ref{thm:intro_match} cannot be strengthened to achieve polynomial-time tractability of $\#\match(\mathcal{G})$ for nowhere dense and monotone $\mathcal{G}$, unless $\ccSharpP=\ccP$.

As a consequence of Theorem~\ref{thm:intro_match} we obtain, for hereditary $\mathcal{H}$, an exhaustive and detailed classification of the complexity of $\#\subsprob(\mathcal{H} \to \mathcal{G})$ as a function of invariants of $\mathcal G$ and $\mathcal H$.

\begin{theorem}\label{thm:subs_hereditary_intro}
Let $\mathcal{H}$ and $\mathcal{G}$ be graph classes such that $\mathcal{H}$ is hereditary and $\mathcal{G}$ is monotone. Then the complexity of $\#\subsprob(\mathcal{H} \to \mathcal{G})$ is exhaustively classified by Table~\ref{tab:resultsSubsHereditary_intro}.
\end{theorem}

\begin{table}[ht]
    \begin{tabularx}{\textwidth}{lllll}
        \toprule
        \begin{tabular}{@{}l}
             $~$
        \end{tabular}&\begin{tabular}{@{}l}
             $\mathcal{G}$
             n.\ dense
        \end{tabular} &\begin{tabular}{@{}l}
             $\mathcal{G}$
             s.\ dense\\
             $\omega(\mathcal{G})=\infty$
        \end{tabular}& \begin{tabular}{@{}l}
             $\mathcal{G}$
             s.\ dense\\
        $\omega(\mathcal{G})<\infty$\\  
            ${\beta}(\mathcal{G})=\infty$
        \end{tabular}&\begin{tabular}{@{}l}
             $\mathcal{G}$
             s.\ dense\\
             $\omega(\mathcal{G})<\infty$\\ 
             ${\beta}(\mathcal{G})<\infty$
        \end{tabular}\\
        \cmidrule(r{2ex}){2-5}
        \begin{tabular}{l}
             $\m(\mathcal{H})<\infty$
        \end{tabular} & P & P & P & P \\[8pt]
        \begin{tabular}{l}
             $\mind(\mathcal{H})= \infty$
        \end{tabular} & FPT & hard & hard & hard
        \\[8pt]
        \begin{tabular}{l}
            $\mind(\mathcal{H})<\infty$,
            $\bind(\mathcal{H})= \infty$
        \end{tabular} & P & hard$^\dagger$ & hard$^\dagger$ & P \\[8pt]
        \begin{tabular}{l}
         Otherwise 
         ~~~
        \end{tabular} & P & hard$^\dagger$ & P & P 
        \\\bottomrule
        \end{tabularx}
     \caption[caption]{\small The complexity of $\#\subsprob(\mathcal{H}\to \mathcal{G})$ for hereditary $\mathcal{H}$ and monotone $\mathcal{G}$. Here ``hard'' means $\#\W[1]$-hard and, unless ETH fails, without an algorithm running in time $f(|H|)\cdot |G|^{o(|V(H)|/ \log |V(H)|)}$; ``hard$^\dagger$'' means the same, but without an algorithm running in $f(|H|)\cdot |G|^{o(|V(H)|)}$.}
      \label{tab:resultsSubsHereditary_intro}
\end{table}
\noindent Note that the unique fixed-parameter tractability result in Table~\ref{tab:resultsSubsHereditary_intro} is a ``real'' $\ccFPT$ case: we can show that, unless $\mathrm{P}=\#\mathrm{P}$, it is in $\ccFPT$ but not in $\ccP$. We point out that the contributions in this work are the hardness results in the third and fourth column, that is, for the cases in which $\calG$ is somewhere dense, but not the class of all graphs. (For monotone $\calG$, $\omega(\calG)=\infty$ implies that $\calG$ is the class of all graphs.)

From the classification of Theorem~\ref{thm:subs_hereditary_intro} one can derive interesting corollaries. For example, when $\mathcal H$ and $\mathcal G$ are monotone one has essentially the same classification of the case $\mathcal G = \mathcal U$: only the boundedness of the matching number of $\scH$ (or equivalently, of its vertex-cover number) counts~\cite{CurticapeanM14}.
\begin{theorem}\label{thm:subs_monotone_intro}
Let $\mathcal{H}$ and $\mathcal{G}$ be monotone classes of graphs and assume that ETH holds. Then $\#\subsprob(\mathcal{H} \to \mathcal{G})$ is fixed-parameter tractable if $\m(\mathcal{H}) < \infty$ or $\mathcal{G}$ is nowhere dense; otherwise $\#\subsprob(\mathcal{H} \to \mathcal{G})$ is $\#\W[1]$-complete and cannot be solved in time $f(|H|)\cdot |G|^{o(|V(H)|/ \log(|V(H)|))}$ for any function $f$.
\end{theorem}

\noindent We conclude by remarking that Table~\ref{tab:resultsSubsHereditary_intro} and the proofs of its bounds suggest the existence of three general algorithmic strategies for subgraph counting:
\begin{enumerate}
    \item If $\scG$ is nowhere dense (first column of Table~\ref{tab:resultsSubsHereditary_intro}), then one can use the FPT algorithm of Theorem~\ref{thm:sparsity_intro}, based on Gaifman's locality theorem for first-order formulas and the local sparsity of nowhere dense graphs (see~\cite{Sparsity}).
    \item If $\m(\scH)< \infty$ (first row of Table~\ref{tab:resultsSubsHereditary_intro}), then one can use the polynomial-time algorithm of Curticapean and Marx~\cite{CurticapeanM14}, based on guessing the image of a maximum matching of $H$ and counting its extensions via dynamic programming.
\item All remaining entries marked as ``P'' are shown to be essentially trivial. Concretely, we will rely on Ramsey's theorem to prove that minor modifications of the naive brute-force approach yield polynomial-time algorithms for those cases.
\end{enumerate}

\subsubsection{The Complexity of $\#\indsubsprob(\scH \to \scG)$}\label{sub:inds}
In the previous section we proved that, when $\scG$ is somewhere dense, $k$-matchings are the minimal hard family of patterns for $\#\subsprob(\scH \to \scG)$. In this section we show that $k$-independent sets play a similar role for $\#\indsubsprob(\scH \to \scG)$. Let $\#\is(\scG)=\#\indsubsprob(\scI \to \scG)$ where $\scI$ is the set of all all independent sets (or $0$-regular graphs). We prove:
\begin{theorem}\label{thm:intro_is}
Let $\mathcal{G}$ be a monotone class of graphs and assume that ETH holds. Then $\#\is(\mathcal{G})$ is fixed-parameter tractable if and only if $\mathcal{G}$ is nowhere dense. More precisely, if $\mathcal{G}$ is nowhere dense then $\#\is(\mathcal{G})$ can be solved in time $f(k)\cdot|V(G)|^{1+o(1)}$ for some computable function $f$; otherwise $\#\is(\mathcal{G})$ cannot be solved in time $f(k)\cdot |G|^{o(k/\log k)}$ for any function $f$. 
\end{theorem}
\noindent This result subsumes the intractability result for counting $k$-independent sets in bipartite graphs of~\cite{CurticapeanDFGL19}. It also strengthens the result of~\cite{BressanR21}, which shows $\#\is(\mathcal{G})$ is hard when parameterized by $k+d$ where $d$ is the degeneracy of $G$. More precisely, \cite{BressanR21} does not imply that $\#\is(\mathcal{G})$ is hard when $\scG$ is the class of $d$-degenerate graphs, for any $d \ge 2$. In contrast to this,  Theorem~\ref{thm:intro_is} proves such hardness for \emph{every} $d \geq 2$. Finally, we point out that the $\ccFPT$ case of Theorem~\ref{thm:intro_is} is not in $\ccP$ unless $\mathrm{P}=\#\mathrm{P}$.

As consequence of Theorem~\ref{thm:intro_is}, when $\scH$ is hereditary (and thus in particular monotone) we obtain:
\begin{theorem}\label{thm:indsubs_hereditary_intro}
Let $\mathcal{H}$ and $\mathcal{G}$ be classes of graphs such that $\mathcal{H}$ is hereditary and $\mathcal{G}$ is monotone. Then the complexity of $\#\indsubsprob(\mathcal{H} \to \mathcal{G})$ is exhaustively classified by Table~\ref{tab:resultsIndSubsHereditary_intro}.
\end{theorem}
\begin{table}[ht]\centering
    \begin{tabularx}{.65\textwidth}{llll}
        \toprule
        \begin{tabular}{@{}l}
             $~$
        \end{tabular}&\begin{tabular}{@{}l}
             $\mathcal{G}$
             n.\ dense
        \end{tabular} &\begin{tabular}{@{}l}
             $\mathcal{G}$
             s.\ dense\\
             $\omega(\mathcal{G})=\infty$
        \end{tabular}& \begin{tabular}{@{}l}
             $\mathcal{G}$
             s.\ dense\\
        $\omega(\mathcal{G})<\infty$\\  
            ${\alpha}(\mathcal{G})=\infty$
        \end{tabular}\\
        \cmidrule(r{2ex}){2-4}
        \begin{tabular}{l}
             $|\mathcal{H}| < \infty$
        \end{tabular} & P & P & P \\[8pt]
        \begin{tabular}{l}
             $\alpha(\mathcal{H})= \infty$
        \end{tabular} & FPT &  hard$^\dagger$ & hard
        \\[8pt]
        \begin{tabular}{l}
           Otherwise 
            ~~
        \end{tabular} & P & hard$^\dagger$ & P\\
        \bottomrule
        \end{tabularx}
     \caption[caption]{\small The complexity of $\#\indsubsprob(\mathcal{H}\to \mathcal{G})$ for hereditary $\mathcal{H}$ and monotone $\mathcal{G}$. Here ``hard'' means $\#\W[1]$-hard and, unless ETH fails, without an algorithm running in time $f(|H|)\cdot |G|^{o(|V(H)|/ \log |V(H)|)}$; ``hard$^\dagger$'' means the same, but without an algorithm running in $f(|H|)\cdot |G|^{o(|V(H)|)}$.}
      \label{tab:resultsIndSubsHereditary_intro}
\end{table}

\subsubsection{The Complexity of $\#\homsprob(\scH \to \scG)$}\label{sub:homs}
Finally, we study the parameterized complexity of $\#\homsprob(\scH \to \scG)$. We denote by $\tw(H)$ the treewidth of a graph $H$. Informally, graphs of small treewidth admit a decomposition with small separators, which allows for efficient dynamic programming. In this work we use treewidth in a purely black-box fashion (e.g.\ via excluded-grid theorems); for its formal definition  see~\cite[Chapter 7]{CyganFKLMPPS15}. We prove:

\vbox{
\begin{theorem}\label{thm:main_homs_intro}
Let $\mathcal{H}$ and $\mathcal{G}$ be monotone classes of graphs. 
\begin{enumerate}
    \item If $\mathcal{G}$ is nowhere dense then $\#\homsprob(\mathcal{H} \to \mathcal{G})$ is fixed-parameter tractable and can be solved in time $f(|H|)\cdot |V(G)|^{1+o(1)}$ for some computable function $f$.
    \item If $\tw(\mathcal{H})< \infty$ then $\#\homsprob(\mathcal{H} \to \mathcal{G})$ is solvable in polynomial time, and if a tree decomposition of $H$ of width $t$ is given, then it can be solved in time $|H|^{O(1)}\cdot |V(G)|^{t+1}$. 
    \item If $\mathcal{G}$ is somewhere dense and $\tw(\mathcal{H})=\infty$ then $\#\homsprob(\mathcal{H} \to \mathcal{G})$ is $\#\W[1]$-hard and, assuming ETH, cannot be solved in time $f(|H|)\cdot |G|^{o(\mathsf{tw}(H))}$ for any function $f$.
\end{enumerate}
(The novel part is {\em 3.}; we included {\em 1}.\ and {\em 2}.\ to provide the complete picture.)
\end{theorem}}

Unfortunately, in contrast to $\#\subsprob$ and $\#\indsubsprob$, we do not know how to extend Theorem~\ref{thm:main_homs_intro} to hereditary $\scH$. We point out however that for hereditary $\mathcal{H}$ the finiteness of $\tw(\scH)$ cannot be the correct criterion: if $\mathcal{H}$ is the set of all complete graphs and $\mathcal{G}$ is the set of all bipartite graphs, then $\scH$ is hereditary and $\tw(\scH)=\infty$, but $\#\homsprob(\mathcal{H} \to \mathcal{G})$ is easy since $|V(H)|\le 2$ or $\#\homs{H}{G}=0$.
More generally, the complexity of $\#\homsprob(\mathcal{H} \to \mathcal{G})$ appears to be far from completely understood for arbitrary classes $\scH$. In fact, it has been recently posed as an open problem even for specific monotone and somewhere dense $\mathcal{G}$ such as the family of $d$-degenerate graphs~\cite{BressanR21,BeraGLSS22}. There is some evidence that the finiteness of induced grid minors is the right criterion for tractability~\cite{BressanR21}.

In what follows we provide a detailed exposition of our proof techniques, starting with a brief summary of the state of the art.

\subsection{Related Work}\label{sec:context}
The general idea of using interpolation as a reduction technique for counting problems dates back to the foundational work of Valiant~\cite{Valiant79b}.
Roughly speaking, the key to interpolation is constructing a system of linear equations that is invertible and thus has a unique solution.
For example, in the classic case of polynomial interpolation (where one has to infer the coefficients of a univariate polynomial given an oracle that evaluates it) the system corresponds to a Vandermonde matrix, which is nonsingular and thus invertible.
In the case of linear combinations of homomorphism counts, an invertible system of linear equations can be constructed via graph tensoring arguments, as proven implicitly by works of Lov{\'{a}}sz (see e.g.\ \cite[Chapters 5 and 6]{Lovasz12}). It was then discovered by Curticapean et al.\ in~\cite{CurticapeanDM17} that these interpolation arguments could be extended to subgraph and induced subgraph counts, by showing that those counts may be expressed as linear combinations of homomorphism counts. Using this fact, they proved that interpolation through graph tensoring applies to a wide variety of parameterised subgraph counting problems. However, their technique fails when one restrict the class of host graphs $\scG$, see the discussion in Section~\ref{sec:concept}; our work shows how to circumvent this obstacle.

The idea of using graph subdivisions for proving hardness results appeared in the context of linear-time subgraph counting in degenerate graphs~\cite{BeraPS20,BeraPS21,BeraGLSS22}. For example,~\cite{BeraPS20} observed that counting triangles in general graphs, which is conjectured not to admit a linear time algorithm, reduces in linear time to counting $6$-cycles in degenerate graphs by subdividing each edge once (which always yields a $2$-degenerate graph). Our work makes heavy use of graph subdivisions as well, although in a more sophisticate fashion. This is not surprising since, for each $d \geq 2$, the class of $d$-degenerate graphs constitutes an example of a monotone somewhere dense class of graphs.

\subsection{Overview of Our Techniques}\label{sec:tech}
The present section expands upon Section~\ref{sec:concept} and gives a detailed technical overview of our proofs of hardness for $\#\subsprob(\mathcal{H}\to\mathcal{G})$ and $\#\indsubsprob(\mathcal{H}\to\mathcal{G})$ (Section~\ref{sub:intro_subind}) and for $\#\homsprob(\mathcal{H}\to\mathcal{G})$ (Section~\ref{sub:intro_hom}). The main contribution of our work is these hardness proofs. The upper bounds hold from (simple adaptations of) previous work. 

\subsubsection{Classifying Subgraph and Induced Subgraph Counting}\label{sub:intro_subind}
We start by analysing a simple case. Recall that a graph family $\mathcal{G}$ is somewhere dense if, for some $r \in \No$, for all $k\in \N$ there is a $G\in \scG$ such that $K_k^r$ is a subgraph of~$G$. From this characterization it is immediate that, if $\scG$ is somewhere dense \emph{and monotone}, then it contains  the $r$-subdivisions of every graph. In turn, this implies that detecting subdivisions of cliques in $\scG$ is at least as hard as the parameterised clique problem~\cite{DvorakKT13}.
Since the parameterised clique problem is $\ccW{1}$-hard, we deduce that $\#\subsprob(\scH \to \scG)$ and $\#\indsubsprob(\scH \to \scG)$ are intractable when $\scH=\{K_k^r : k,r \in \N\}$ and $\scG$ is monotone and somewhere dense. 
Unfortunately, it is unclear how to extend this approach to arbitrary $\scH$, since the elements of $\scH$ are not necessarily $r$-subdivisions of graphs that are hard to count. To show how this obstacle can be overcome, we will focus on $\#\subsprob(\scH \to \scG)$ when $\scH$ is the class of $k$-matchings, $\mathcal{M} = \{M_k : k \in \N\}$; in other words, on the problem of counting $k$-matchings, $\#\subsprob(\mathcal M \to \scG)$. This problem will turn out to be the minimal hard case for $\#\subsprob(\scH \to \scG)$, and its analysis will contain the key ingredients of our proof. The proof for $\#\indsubsprob(\scH \to \scG)$ will be similar.

Let us start by outlining the hardness proof of $\#\subsprob(\mathcal{M} \to \scG)$ when $\scG=\scU$, by using the interpolation technique discussed in Section~\ref{sec:context}. From~\cite{CurticapeanDM17}, we know that for every $k \in \N$ there is a function $a_k : \all \to \Q$ with finite support such that, for every $G \in \all$,
\begin{equation}\label{eq:intro_hombasis}
    \#\subs{M_k}{G} = \sum_H a_k(H) \cdot \#\homs{H}{G}
\end{equation}
where the sum is over all isomorphism classes of all graphs. By a classic result of Dalmau and Jonsson~\cite{DalmauJ04}, computing $\#\homs{H}{G}$ is not fixed-parameter tractable for $H$ of unbounded treewidth, unless ETH fails. Hence, if we could use~\eqref{eq:intro_hombasis} to show that an FPT algorithm for computing $\#\subs{M_k}{G}$ yields an FPT algorithm for computing $\#\homs{H}{G}$ for some $H$ whose treewidth grows with $k$, we would conclude that computing $\#\subs{M_k}{G}$ is not fixed-parameter tractable unless ETH fails. 
This is what~\cite{CurticapeanDM17} indeed prove.
The idea is to apply~\eqref{eq:intro_hombasis} not to $G$, but to a set of carefully chosen graphs
$\hat{G}_1,\ldots,\hat{G}_{\ell}$ such that the counts $\#\homs{M_k}{\hat{G}_1},\ldots,\#\homs{M_k}{\hat{G}_{\ell}}$ can be used to solve a linear system and infer $\#\homs{H}{G}$ for all $H$ appearing on the right-hand side of~\eqref{eq:intro_hombasis}. 

Let us explain this idea in more detail. Suppose we had an oracle for $\#\subsprob(\mathcal M \to \all)$, so that we could quickly compute $\#\subs{M_k}{G}$ for any desired $G$. 
Let $\ell$ be the size of the support of $a_k$, which is finite and thus a function of $k$, and let $\{G_i\}_{i=1,\ldots,\ell}$ be a set of graphs such that each $G_i$ has size bounded by a function of $k$.
It is a well-known fact that, for all graphs $H, G, G'$,
\begin{equation}\label{eq:intro_tensor_linear}
\#\homs{H}{G\times G'} = \#\homs{H}{G}\cdot \#\homs{H}{G'}.
\end{equation}
By combining~\eqref{eq:intro_hombasis} and~\eqref{eq:intro_tensor_linear}, for each $i=1,\ldots,\ell$ we obtain 
\begin{align}\label{eq:intro_hombasis_tensor}
    \#\subs{M_k}{G\times G_i} &=
    \sum_{H} a_{k}(H) \cdot \#\homs{H}{G_i} \cdot \#\homs{H}{G} = \sum_{\substack{H\\a_k(H)\neq 0}} b^i_H \cdot X_H\,,
\end{align}
where $b^i_H := \#\homs{H}{G_i}$ and $X_H:=a_{k}(H) \cdot \#\homs{H}{G}$.
Now, we can compute $\#\homs{H}{G_i}$ in FPT time since $|G_i|$ is bounded by a function of $k$, and we can compute $\#\subs{M_k}{G\times G_i}$ using the oracle. Therefore, in FPT time we can comput a system of $\ell$ linear equations with the $X_H$ as unknowns. By applying classical results due to Lov{\'{a}}sz (see e.g.\ \cite[Chapter 5]{Lovasz12}), Curticapean et al.\ \cite{CurticapeanDM17} showed that there always exists a choice of the $G_i$'s such that this system has a unique solution. Hence, using those $G_i$'s one can compute $\#\homs{H}{G}$ in FPT time for all $H$ with $a_{k}(H) \ne 0$. In particular, one can compute $\#\homs{F_k}{G}$ where $F_k$ is any $k$-edge graph of maximal treewidth, since~\cite{CurticapeanDM17} also showed that $a_{k}(H) \ne 0$ for all $H$ with $|E(H)| \le k$.
This gives a parameterized reduction from $\#\homsprob(\mathcal{F} \to \all)$ to $\#\subsprob(\mathcal{M} \to \all)$, where $\mathcal{F}$ is the class of all maximal-treewidth graphs $F_k$. Since $\#\homsprob(\mathcal{F} \to \all)$ is hard by~\cite{DalmauJ04}, the reduction establishes hardness of $\#\subsprob(\mathcal{M} \to \all)$ as desired.

Our main question is whether this strategy can be extended from $\scU$ to any monotone somewhere dense class $\scG$. This it not obvious, since the argument above relies on two crucial ingredients that may be lost when moving from $\scU$ to $\scG$:
\begin{enumerate}
    \item[(I.1)] We need to find a family of graphs $\hat{\mathcal{F}}= \{ \hat{F}_k~|~k\in \N\}$ such that $\#\homsprob(\hat{\mathcal{F}} \to \scG)$ is hard
    and, for all $k\in \N$,
    $a_k(\hat{F}_k) \neq 0$. 
    \item[(I.2)] We need to find graphs $G_i$ such that $G \times G_i \in \scG$. This is necessary since the argument performs a reduction to the problem of counting $\#\subs{M_k}{G\times G_i}$, and is not straightforward
    since $G\times G_i$ may not be in $\scG$ even when both $G,G_i$ are.
\end{enumerate}
It turns out that both requirements can be satisfied in a systematic way.
First, we study $\#\subsprob(\mathcal{H}\to\mathcal{G})$ in some carefully chosen vertex-coloured and edge-coloured version. It is well-known that the coloured version of the problem is equivalent in complexity (in the FPT sense) to the uncoloured version; so, to make progress, we may consider the coloured version. Next, coloured graphs come with a canonical coloured version of the tensor product which satisfies (\ref{eq:intro_tensor_linear}), so we can hope to apply interpolation via tensor products in the colorful setting, too. The introduction of colours in the analysis of 
parameterised problems is a common tool for streamlining reductions that are otherwise unnecessarily complicated (see e.g.\ \cite{CurticapeanM14,PeyerimhoffRSSVW22,DorflerRSW22,FockeR22}). The technical details of the coloured version are not hard, but cumbersome to state; since here we do not need them, we defer them to Section~\ref{sec:prelims}. Let us now give a high-level explanation of how we achieve (I.1) and (I.2).

For (I.1), we let $\hat{\mathcal{F}}$ be the class of all $r$-subdivisions of a family $\mathcal{E}$ of regular expander graphs. A simple construction then allows us to reduce $\#\homsprob(\mathcal{E} \to \all)$, which is known to be hard, to $\#\homsprob(\hat{\mathcal{F}}\to \all^r)$, where $\all^r$ is the set of all $r$-subdivisions of graphs. As noted above $\all^r \subseteq \mathcal{G}$, hence 
$\#\homsprob(\hat{\mathcal{F}} \to \mathcal{G})$ is hard.
We will show in the coloured version that for each graph $F_k \in \hat{\mathcal{F}}$ with $k$ edges, $a_k(F_k)\neq 0$ (see the proof of Lemma~\ref{lem:main_colmatch}). Thus, (I.1) is satisfied.

For (I.2) we construct, for each $k$, a finite sequence of coloured graphs $G_1, G_2,\dots$ satisfying the following two conditions: the system of linear equations given by (the coloured version of)~\eqref{eq:intro_hombasis_tensor} has a unique solution, and the coloured tensor product between each $G_i$ and any coloured graph in $\all^r$ is in $\mathcal{G}$. Concretely, we choose as $G_i$ the so-called \emph{fractured graphs} of the $r$-subdivisions of the expanders in $\mathcal{E}$. Fractured graphs are obtained by a splitting operation on a graph and come with a natural vertex colouring. They have been introduced in recent work on classifying subgraph counting problems~\cite{PeyerimhoffRSSVW22} and we describe them in Section~\ref{sec:prelims_fractures}.

Together, our resolutions of (I.1) and (I.2) yield a colourful version of the framework of~\cite{CurticapeanDM17} that applies to \emph{any monotone somewhere dense class of host graphs}. As a consequence we obtain that  $\#\homsprob(\mathcal{E} \to \all)$, the problem of counting homomorphisms from expanders in $\mathcal{E}$ to arbitrary hosts graphs, reduces in FPT time to $\#\subsprob(\mathcal{M} \to \mathcal{G})$ whenever $\mathcal{G}$ is monotone and somewhere dense. Since $\#\homsprob(\mathcal{E} \to \all)$ is intractable, this proves the hardness of $\#\subsprob(\mathcal{M} \to \mathcal{G})$ for all monotone and somewhere dense $\scG$, as stated in Theorem~\ref{thm:intro_match}. From this result we will then be able to prove our general classification for $\#\subsprob(\mathcal{H}\to \mathcal{G})$ (Theorem~\ref{thm:subs_hereditary_intro}) by combining existing results and Ramsey-type arguments on $\mathcal H$ and $\mathcal G$.

This concludes our overview for $\#\subsprob(\mathcal H \to \mathcal G)$. The proofs for $\#\indsubsprob(\mathcal H \to \mathcal G)$ are similar, but instead of $\#\subsprob(\mathcal{M} \to \mathcal{G})$, they use as a minimal hard case $\#\is(\mathcal{G})$, the problem of counting $k$-independent sets in host graphs from $\mathcal G$.

\subsubsection{Classifying Homomorphism Counting via Wall Minors}\label{sub:intro_hom}
The proof of our dichotomy for $\#\homsprob(\mathcal{H} \to \mathcal{G})$ for monotone $\mathcal{H}$ and $\mathcal{G}$ (Theorem~\ref{thm:main_homs_intro}) requires us to establish hardness when $\mathcal{G}$ is somewhere dense and $\tw(\mathcal{H})=\infty$. Recall that our solution of (I.1) relied on a reduction from (the coloured version of) $\#\homsprob(\mathcal{E}\to \all)$ to (the coloured version of) $\#\homsprob(\hat{\mathcal{F}} \to \all^r)$, where $\mathcal{E}$ is a family of regular expander graphs, $\hat{\mathcal{F}}$ is the class of all $r$-subdivisions of graphs in $\mathcal{E}$, and $\all^r$ is the class of $r$-subdivisions of all graphs. Since for all monotone somewhere dense classes $\mathcal{G}$ there is an $r$ such that $\all^r\subseteq \mathcal{G}$, we would be done if we could make sure that every monotone class of graphs of unbounded treewidth $\mathcal{H}$ contains $\hat{\mathcal{F}}$ as a subset. Unfortunately, this is not the case. As a trivial example, $\mathcal{H}$ could be the class of all graphs of degree at most $3$ while $\mathcal{E}$ is a family of $4$-regular expanders.

To circumvent this problem, we use a result of Thomassen~\cite{Thomassen88} to prove that, for every positive integer $r$ and every monotone class of graphs $\mathcal{H}$ with unbounded treewidth, the following holds: for each \emph{wall} $W_{k,k}$, the class $\mathcal{H}$ contains a subdivision of $W_{k,k}$ in which each edge is subdivided a positive multiple of $r$ times. Now, the crucial property of the class of all walls $\mathcal{W}:=\{W_{k,k}~|~k\in \N\}$ is that $\#\homsprob(\mathcal{W}\to \mathcal{U})$ is intractable by the classification of Dalmau and Jonsson~\cite{DalmauJ04}. Refining our constructions based on subdivided graphs, we are then able to show that $\#\homsprob(\mathcal{W}\to \mathcal{U})$ reduces to $\#\homsprob(\mathcal{H}\to\mathcal{G})$ whenever $\mathcal{H}$ is monotone and of unbounded treewidth, and $\mathcal{G}$ is monotone and somewhere dense. Theorem~\ref{thm:main_homs_intro} will then follow as a direct consequence.

\section{Preliminaries}\label{sec:prelims}
We denote the set of non-negative integers by $\No$, and the set of positive integers by $\N$.
Graphs in this work are undirected and without self-loops unless stated otherwise. A \emph{subdivision} of a graph $G$ is obtained by subdividing each edge of $G$ arbitrarily often. Given a graph $G$ and $r\in\No$, we write $G^r$ for the $r$-\emph{subdivision} of $G$, i.e., the graph obtained from $G$ by subdividing each edge   $r$ times (so that it becomes a path of $r+1$ edges). Note that $G^0=G$. (The graph~$G^{r-1}$ is also called the ``$r$-stretch of~$G$'' in the literature).   Given a graph $G$ and a vertex $v\in V(G)$, we write $E_G(v):= \{e\in E(G)~|~v\in e \}$ for the set of edges incident to $v$. Furthermore, given $A\subseteq E(G)$, we write $G[A]$ for the 
graph~$(V(G),A)$.
Given a subset of vertices $S\subseteq V(G)$, we write $G[S]$ for the subgraph of $G$ induced by the vertices in $S$, that is, 
$G[S] := (S, \{e\in E(G)~|~e \subseteq S \})$. An ``induced subgraph'' of~$G$ is a subgraph induced by some $S\subseteq V(G)$.

A \emph{homomorphism} from a graph $H$ to a graph $G$ is a mapping $\varphi:V(H)\to V(G)$ which is edge-preserving, that is, $\{u,v\}\in E(H)$ implies $\{\varphi(u),\varphi(v)\}\in E(G)$. 
We write:
\begin{itemize}\itemsep0pt
\item $\homs{H}{G}$ for the set of all homomorphisms from $H$ to $G$,
\item $\surhoms{H}{G}$ for the set of all surjective homomorphisms
from~$H$ to~$G$,
\item $\subs{H}{G}$ for the set of all subgraphs of $G$ isomorphic to $H$, and
\item $\indsubs{H}{G}$ for the set of all induced subgraphs of $G$ isomorphic to $H$.
\end{itemize}

\subsection{Coloured Graphs and Fractures}\label{sec:prelims_fractures}
Let $H$ be a graph. Following standard terminology, we refer to an element of $\homs{G}{H}$ as an \emph{$H$-colouring} of the graph~$G$. An $H$-\emph{coloured graph} is a pair $(G,c)$ where $G$ is a graph and $c$ an $H$-colouring of $G$. 
We say that $(G,c)$ is a surjectively $H$-coloured graph if  $c \in \surhoms{G}{H}$.

Given two $H$-coloured graphs $(F,c_F)$ and $(G,c_G)$, a homomorphism from $(F,c_F)$ to $(G,c_G)$ is a mapping $\varphi\in\homs{F}{G}$ such that
$c_G(\varphi(v))=c_F(v)$ for each $v\in V(F)$.\footnote{We remark that in previous work~\cite{PeyerimhoffRSSVW22}, homomorphisms between $H$-coloured graphs are called ``colour-preserving'' or, if $F=H$, ``colour-prescribed''. Since we will work almost exclusively in the coloured setting in this work, we will just speak of homomorphisms and always provide the $H$-colourings explicitly in our notation.} We write $\homs{(F,c_F)}{(G,c_G)}$ for the set of all homomorphisms from $(F,c_F)$ to $(G,c_G)$.

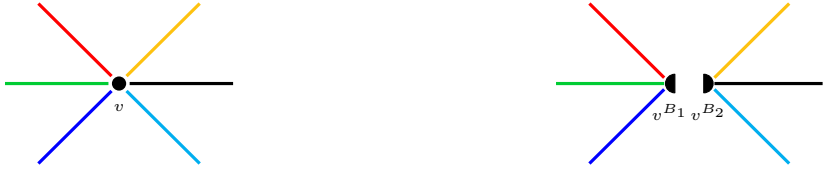
\begin{figure}[t]
    \centering
    \begin{tikzpicture}[scale=1.5]
        \node[vertex,inner sep=.4ex,label={[label distance=.03]below:\(v\)}] (m) at (0, 0) {};

        \draw[very thick,red] (m) -- ++(135:1);
        
        \draw[very thick,green!80!blue] (m) -- ++(180:1);
        
        \draw[very thick,blue] (m) -- ++(-135:1);

        \draw[very thick,yellow!50!orange] (m) -- ++(45:1);
        
        \draw[very thick] (m) -- ++(0:1);
        
        \draw[very thick,cyan] (m) -- ++(-45:1);

        \begin{scope}[shift={(5,0)}]
            
            \begin{scope}[scale=2.4]
                \kowaen{0,0}{-90/90/white,90/270/white}{1};
            \end{scope}

            \node[label={[label distance=.03]below:\(v^{B_1}\)}]  at (1-2) {};

            \draw[very thick,red] (1-2) -- ++(135:1);
            
            \draw[very thick,green!80!blue] (1-2) -- ++(180:1);
            
            \draw[very thick,blue] (1-2) -- ++(-135:1);

            \node[label={[label distance=.03]below:\(v^{B_2}\)}]  at
                (1-1) {};

            \draw[very thick,yellow!50!orange] (1-1) -- ++(45:1);
            
            \draw[very thick] (1-1) -- ++(0:1);
            
            \draw[very thick,cyan] (1-1) -- ++(-45:1);

            \begin{scope}[scale=1.8]
                \kowaen{0,0}{-90/90/black,90/270/black}{1};
            \end{scope}

        \end{scope}
    \end{tikzpicture}
    \caption{\label{fig:fracture} A fractured graph $\fracture{Q}{\sigma}$ from~\cite{PeyerimhoffRSSVW22}. Left: a vertex $v \in V(Q)$ with incident edges $E_Q(v)=\{
        \redc{2pt},\greenc{2pt},\bluec{2pt},\yellowc{2pt},\blackc{2pt},\brownc{2pt}\}$. Right: the splitting of $v$ in $\fracture{Q}{\sigma}$ for a fracture $\sigma$ where the partition $\sigma_v$ of $E_Q(v)$ consists of the two blocks $B_1 =\{ \redc{2pt},\greenc{2pt},\bluec{2pt}\}$, and $B_2=\{\yellowc{2pt},\blackc{2pt},\brownc{2pt}\}$.}
\end{figure}

Following the terminology of~\cite{PeyerimhoffRSSVW22}, we define a fracture of a graph $H$ as a $|V(H)|$-tuple $\rho=(\rho_v)_{v\in V(H)}$ where $\rho_v$ is a partition of the set $E_H(v)$ of edges of $H$ incident to $v$. Now, given a fracture $\rho$ of $H$, we obtain the \emph{fractured graph} $\fracture{H}{\rho}$ from $H$ by splitting each vertex $v$ according to the partition $\rho_v$. Formally, the graph $\fracture{H}{\rho}$ contains a vertex $v^B$ for each vertex $v\in V(H)$ and block $B \in \rho_v$, and we make $v^B$ and $u^{B'}$ adjacent if and only if $\{v,u\}\in E(H)$ and $\{u,v\} \in B \cap B'$. An illustration is provided in Figure~\ref{fig:fracture}.

The following $H$-colouring of a fractured graph is used implicitly in~\cite{PeyerimhoffRSSVW22}. 
\begin{definition}\label{def:canonical_col}
Let $H$ be a graph and $\rho$ a fracture of $H$. We denote by $c_\rho:V(\fracture{H}{\rho}) \to V(H)$ the function that maps $v^B$ to $v$ for each $v\in V(H)$ and $B\in \rho_v$.
\end{definition}
\begin{observation}\label{obs:canonical_col}
For each $H$ and $\rho$,   $c_\rho$ is an $H$-colouring of the fractured graph $\fracture{H}{\rho}$.
\end{observation}

\subsection{Graph Classes, Invariants and Minors}\label{sub:prelim_invariants}
We use symbols $\mathcal{F},\mathcal{G},\mathcal{H}$ to denote classes of graphs, and we denote by $\all$ be the class of \emph{all} graphs. A graph invariant is a function $g : \all \to \No$ such that $g(G)=g(H)$ whenever $G$ and $H$ are isomorphic. An invariant $g$ is \emph{bounded} on a graph family $\mathcal{H}$ if there exists $B \in \No$ such that $g(H)\leq B$ for all $H\in \mathcal{H}$, in which case we write $g(\mathcal H)<\infty$; otherwise we say $g$ is unbounded on $\mathcal{H}$ and write $g(\mathcal H)=\infty$. Our statements involve the following invariants.
\begin{definition}[Graph Invariants]\label{def:invariants}
For any graph $G$ define:
\begin{itemize}\itemsep0pt
    \item the independence number $\alpha(G)$, i.e., the size of the largest independent set of $G$
    \item the clique number $\omega(G)$, i.e., the size of the largest complete subgraph of $G$
    \item the biclique number $\beta(G)$, i.e., the largest $k$ such that $G$ contains a $k$-by-$k$ biclique as a subgraph, and its induced version, the induced biclique number $\bind(G)$
    \item the matching number $\m(G)$, i.e., the size of a maximum matching of $G$, and its induced version, the induced matching number $\mind(G)$
\end{itemize}
\end{definition}
We denote by $\mathsf{tw}(G)$ the \emph{treewidth} of a graph $G$. We omit the definition of treewidth as we rely on it in a black-box manner; the interested reader can see e.g.\ Chapter 7 of~\cite{CyganFKLMPPS15}.
For any $k \in \N$ the $k$-by-$k$ grid graph $\boxplus_k$, depicted in Figure~\ref{fig:wall_and_grid}, is defined by $V(\boxplus_k)=[k]^2$ and $E(\boxplus_k)=\{\{(i,j),(i',j')\} : i,j,i',j' \in [k], |i-i'|+|j-j'|=1\}$.
It is well known that $\tw(\boxplus_k)=k$, see~\cite[Chapter 7.7.1]{CyganFKLMPPS15}.

We make use of the following two consequences of Ramsey's Theorem for an arbitrary class of graphs $\scH$. The first one is immediate, and the second one was established by Curticapean and Marx in~\cite{CurticapeanM14}.
\begin{theorem}\label{thm:ramsey}
If $|\mathcal{H}|=\infty$ then $\max(\alpha(\scH),\omega(\scH))=\infty$.
\end{theorem}
\begin{theorem}\label{thm:ramsey_match}
If $\m(\mathcal{H})=\infty$ then $\max(\omega(\scH),\bind(\scH),\mind(\scH))=\infty$.
\end{theorem}

A class of graphs is \emph{hereditary} if it is closed under vertex deletion, and is \emph{monotone} if it is hereditary and closed under edge deletion. In other words, hereditary classes are closed under taking induced subgraphs, and monotone classes are closed under taking subgraphs.

To present the different notions of graph minors used in this paper in a unified way, we start by introducing \emph{contraction models}.

\begin{definition}[Contraction model]
A \emph{contraction model} of a graph $H$ in a graph $G$ is a partition $\{V_1,\dots,V_k\}$ of $V(G)$ such that $G[V_i]$ is connected for each $i\in[k]$ and that $H$ is isomorphic to the graph obtained from $G$ by contracting each $G[V_i]$ into a single vertex (and deleting multiple edges and self-loops).
\end{definition}

Recall that a graph $F$ is a \emph{minor} of a graph $G$ if $F$ can be obtained from $G$ by deletion of edges and vertices, and by contraction of edges; equivalently, $F$ is a minor of $G$ if $F$ is a subgraph of a graph that has a contraction model in $G$. In this work, we will also require the subsequent stricter notion of minors.

\begin{definition}[Shallow minor~\cite{NesetrildM11}]
A graph $F$ is a \emph{shallow minor} at depth $d$ of a graph $G$ if $F$ is a subgraph of graph $H$ that has a contraction model $\{V_1,\dots,V_k\}$ in $G$ satisfying the following additional constraint: for each $i\in[k]$ there is a vertex $x_i\in V_i$ such that each vertex in $V_i$ has distance at most $d$ from $x_i$.
Given a class of graphs $\mathcal{G}$, we write $\mathcal{G} \triangledown d$ for the set of all shallow minors at depth $d$ of graphs in $\mathcal{G}$.
\end{definition}

Observe that the shallow minors at depth $0$ of $G$ are exactly the subgraphs of $G$, and the shallow minor of depth $|V(G)|$ are exactly the minors of $G$. For this reason, the notion of a shallow minor can be considered an interpolation between subgraphs and minors. Furthermore, having introduced this notion, we are now able to define somewhere dense and nowhere dense graph classes. 

\begin{definition}[Somewhere dense graph classes~\cite{NesetrildM11}]
A class of graphs $\mathcal{G}$ is \emph{somewhere dense} if $\omega(\mathcal{G} \triangledown d)=\infty$ for some $d \in \No$, and is \emph{nowhere dense} if instead $\omega(\mathcal{G} \triangledown d)< \infty$ for all $d \in \No$.
\end{definition}

We use the following characterisation of monotone somewhere dense graph classes.\footnote{It is non-trivial to pinpoint the first statement of Lemma~\ref{lem:nd_subdivisions} in the literature: Dvor{\'{a}}k et al.\ \cite{DvorakKT13} attribute it to Ne{\v{s}}et{\v{r}}il and de Mendez~\cite{NesetrildM11}, who provide an implicit proof. The first explicit statement is, to the best of our knowledge, due to Adler and Adler~\cite{Adlers14}.}

\begin{lemma}[Remark~2 in~\cite{Adlers14}]\label{lem:nd_subdivisions}
Let $\mathcal{G}$ be a monotone class of graphs. Then $\mathcal{G}$ is somewhere dense if and only if there exists $r \in \No$ such that $G^r \in \mathcal{G}$ for all $G\in \all$.
\end{lemma}

\subsection{Parameterised and Fine-Grained Complexity}\label{sec:param_fg}

A \emph{parameterized counting problem} is a pair $(P,\kappa)$ where $P:\{0,1\}^\ast \rightarrow \N$ and $\kappa:\{0,1 \}^\ast \rightarrow \N$ is computable. For an instance $x$ of $P$ we call $\kappa(x)$ the \emph{parameter} of $x$.
An algorithm $\mathbb{A}$ is \emph{fixed-parameter tractable} (FPT) w.r.t.\ a parameterization $\kappa$ if there is a computable function $f$ such that $\mathbb{A}$ runs in time $f(\kappa(x))\cdot |x|^{O(1)}$ on every input $x$. A parameterized counting problem $(P,\kappa)$ is \emph{fixed-parameter tractable} (FPT) if there is an FPT algorithm (w.r.t.\ $\kappa$) that computes $P$.

A \emph{parameterized Turing reduction} from $(P,\kappa)$ to $(P',\kappa')$ is an algorithm $\mathbb{A}$ equipped with oracle access to $P'$ satisfying the following constraints:
\begin{itemize}\itemsep0pt
    \item[(A1)] $\mathbb{A}$ computes $P$
    \item[(A2)] $\mathbb{A}$ is FPT w.r.t.\ $\kappa$
    \item[(A3)] there is a computable function $g$ such that, on input $x$, each oracle query $x'$ satisfies that $\kappa'(x')\leq g(\kappa(x))$.
\end{itemize}
We write $(P,\kappa) \fptred (P',\kappa')$ if a parameterized Turing reduction from $(P,\kappa)$ to $(P',\kappa')$ exists. 

The parameterized counting problem $\#\clique$ asks, on input a graph $G$ and $k \in \N$, to compute the number of $k$-cliques in $G$; the parameter is $k$. As shown by Flum and Grohe~\cite{FlumG04}, $\#\clique$ is the canonical complete problem for the parameterized complexity class $\#\W[1]$. In particular, a parameterized counting problem $(P,\kappa)$ is called $\#\W[1]$-\emph{hard} if $\#\clique \fptred (P,\kappa)$. We omit the technical definition of $\#\W[1]$ via weft-1 circuits (see Chapter 14 of~\cite{FlumG06}), but we recall that $\#\W[1]$-hard problems are not FPT unless standard hardness assumptions fail (see below).

We define the problems studied in this work. As usual $\mathcal{H}$ and $\mathcal{G}$ denote classes of graphs. 

\begin{definition}
$\#\homsprob(\mathcal{H}\!\to\!\mathcal{G}),\#\subsprob(\mathcal{H} \to \mathcal{G}),\#\indsubsprob(\mathcal{H} \to \mathcal{G})$ ask, given $H \in \mathcal{H}$ and $G\in \mathcal{G}$, to compute respectively $\#\homs{H}{G},\#\subs{H}{G}, \#\indsubs{H}{G}$. The parameter is $|H|$.
\end{definition}

\noindent For example, $\#\subsprob(\mathcal{H}\to \mathcal{G})=\#\clique$ when $\mathcal{H}$ is the class of all complete graphs and $\mathcal{G}$ the class of all graphs. The following result follows immediately from an algorithm for counting answers to Boolean queries in nowhere dense graphs due to Ne{\v{s}}et{\v{r}}il and Ossona de Mendez~\cite{Sparsity}.

\begin{theorem}[Theorem 18.9 in~\cite{Sparsity}]\label{thm:sparsity}
If $\mathcal{G}$ is nowhere dense then $\#\homsprob(\mathcal{H} \to \mathcal{G})$, $\#\subsprob(\mathcal{H} \to \mathcal{G})$, and $\#\indsubsprob(\mathcal{H} \to \mathcal{G})$ are fixed-parameter tractable and can be solved in time
$f(|H|)\cdot |V(G)|^{1+o(1)}$ for some computable function $f$.
\end{theorem}

In an intermediate step towards our classifications, we will rely on a coloured version of homomorphism counting.
\begin{definition}\label{defwithsurj}
$\#\cphomsprob(\mathcal{H} \to \mathcal{G})$ asks, given $H\in \mathcal{H}$ and a surjectively\footnote{In previous works (e.g.\ in~\cite{PeyerimhoffRSSVW22}), the definition of $\#\cphomsprob(\mathcal{H} \to \mathcal{G})$ did not require the $H$-colouring to be surjective. However, one can always assume surjectivity, since $\#\homs{(H,\id_H)}{(G,c)}=0$ if $c$ is not surjective. We decided to make the surjectivity condition explicit in the current work.} $H$-coloured graph $(G,c)$ with $G \in \mathcal G$, to compute $\#\homs{(H,\id_H)}{(G,c)}$, where $\id_H$ denotes the identity on $V(H)$. The parameter is $|H|$.
\end{definition}
\noindent It is well known that $\#\cphomsprob(\mathcal{H} \to \all)$ reduces to the uncoloured version via inclusion-exclusion. The same holds for $\#\cphomsprob(\mathcal{H} \to \mathcal{G})$, too, if $\mathcal{G}$ is monotone. Formally:
\begin{lemma}[see e.g.\ Lemma~2.49 in~\cite{Roth19}]\label{lem:hom_colreduction}
If $\mathcal{G}$ is monotone then $\#\cphomsprob(\mathcal{H} \to \mathcal{G}) \fptred \#\homsprob(\mathcal{H}\to \mathcal{G})$.
Moreover, on input $H\in \mathcal{H}$ and $(G,c)$ with $G\in \mathcal{G}$, every oracle query $(H',G')$ in the reduction satisfies $H'=H$ and $G' \subseteq G$.
\end{lemma}
\noindent An implicit consequence of the parameterized complexity classification for counting homomorphisms due to Dalmau and Jonsson~\cite{DalmauJ04} establishes the following hardness result for $\#\cphomsprob$; an explicit argument can be found e.g.\ in Chapter~2 in~\cite{Roth19}.
\begin{theorem}[\cite{DalmauJ04}]\label{thm:cphomshard}
If $\scH$ is recursively enumerable and $\tw(\scH)=\infty$ then $\#\cphomsprob(\mathcal{H} \to \all)$ is $\#\W[1]$-hard.
\end{theorem}

Finally, all running-time lower bounds in this paper are conditional on ETH:
\begin{definition}[\cite{ImpagliazzoP01}]
The \emph{Exponential Time Hypothesis (ETH)} asserts that $3$-$\textsc{SAT}$ cannot be solved in time $\mathsf{exp}(o(n))$ where $n$ is the number of variables of the input formula. 
\end{definition}
\noindent Chen et al.~\cite{Chenetal05,Chenetal06} showed that there is no function $f$ such that $\#\clique$ can be solved in time $f(k)\cdot |G|^{o(k)}$ unless ETH fails. This in particular implies that $\#\W[1]$-hard problems are not FPT unless ETH fails. Marx~\cite{Marx10} strengthened Theorem~\ref{thm:cphomshard} into:\footnote{More precisely, Marx established the bound for the so-called partitioned subgraph problem. However, as shown in~\cite{RothSW20}, the lower bound immediately translates to $\#\cphomsprob(\mathcal{H}\to \all)$.}

\begin{theorem}[\cite{Marx10}]\label{thm:cybt}
If $\mathcal{H}$ is recursively enumerable and $\tw(\mathcal{H})=\infty$ then $\#\cphomsprob(\mathcal{H}\to \all)$ cannot be solved in time $f(|H|)\cdot |G|^{o\left(\frac{\mathsf{tw}(H)}{\log \mathsf{tw}(H)}\right)}$
for any function $f$, unless ETH fails.
\end{theorem}
\noindent The question of whether the $(\log \mathsf{tw}(H))^{-1}$ factor in the above lower bound can be omitted can be considered the counting version of the open problem ``Can you beat treewidth?''~\cite{Marx10,Marx13}.

\section{Counting Homomorphisms}
This section is devoted to the proof of our dichotomy theorem for $\#\homsprob(\mathcal{H} \to \mathcal{G})$, Theorem~\ref{thm:main_homs_intro}. We start by showing a reduction from $\#\cphomsprob(\mathcal{H}\to \all)$ to counting colour-prescribed homomorphisms between subdivided graphs. While the proof is straightforward, the reduction will turn out useful for the more involved cases of $\#\subsprob(\mathcal{H} \to \mathcal{G})$ and $\#\indsubsprob(\mathcal{H} \to \mathcal{G})$. Theorem~\ref{thm:main_homs_intro} will be an immediate consequence of Theorem~\ref{thm:sparsity} and Theorem~\ref{thm:main_homs} below.

To begin with, let 
$c \in \surhoms{G}{H}$ and let $r \in \No$. Define the following canonical
homomorphism $c^r$ from~$G^r$ to~$H^r$.  For each $u\in V(G)$, set $c^r(u):= c(u)$. 
For any edge $e=\{u_1,u_2\} \in E(G)$,  let $u_1,w_1,\dots,w_r,u_2$ be the corresponding path in $G^r$. Let $e'=\{v_1,v_2\}=\{c(u_1),c(u_2)\}$ --- note that $e' \in E(H)$ as $c \in \homs{G}{H}$ --- and let $v_1,x_1,\dots,x_r,v_2$ be the corresponding path in $H^r$. Then, set $c^r(w_i):=x_i$ for each $i\in\{1,\dots,r\}$.
It is easy to see that $c^r$ is a surjective $H^r$-colouring of $G^r$. Furthermore:
\begin{lemma}\label{lem:hom_subdiv}\label{lem:aa}
For every surjectively $H$-coloured graph $(G,c)$ and every $r \in \No$,
\begin{align}
    \#\homs{(H,\id_H)}{(G,c)}=\#\homs{(H^r,\id_{H^r})}{(G^r,c^r)}
\end{align}
where $\id_H$ and $\id_{H^r}$ are the identities on respectively  $V(H)$ and $V(H^r)$.
\end{lemma}
\begin{proof}
 We define a bijection $b:\homs{(H,\id_H)}{(G,c)} \to \homs{(H^r,\id_{H^r})}{(G^r,c^r)}$. Let $\varphi \in \homs{(H,\id_H)}{(G,c)}$.
For every $v\in V(H)$ let $b(\varphi)(v)=\varphi(v)$. For every $\{v_1,v_2\} \in E(H)$ and every $i \in [r]$, if $u_1=\varphi(v_1)$ and $u_2=\varphi(v_2)$, and if $x_i$ and $w_i$ are the $i$-th vertices on the paths respectively between $v_1$ and $v_2$ in $H^r$ and between $u_1$ and $u_2$ in $G^r$, then let $b(\varphi)(x_i)=w_i$.
It is easy to see that $b(\varphi) \in \homs{(H^r,\id_{H^r})}{(G^r,c^r)}$ and that $b$ is injective. To see that $b$ is surjective as well, note that for every $\varphi^r \in \homs{(H^r,\id_{H^r})}{(G^r,c^r)}$ its restriction $\varphi^r|_{V(H)}$ to $V(H)$ satisfies $\varphi^r|_{V(H)} \in \homs{(H,\id_H)}{(G,c)}$ and $b(\varphi^r|_{V(H)})= \varphi^r$.
\end{proof}

\subsection{Warm-up: Minor-closed Pattern Classes}
Using the characterisation of somewhere dense graph classes in Lemma~\ref{lem:nd_subdivisions}, and known lower bounds for counting homomorphisms from grid graphs, we obtain as an easy consequence the following complexity dichotomy:
\begin{theorem}\label{thm:homs_warmup}
Let $\mathcal{H}$ be a minor-closed class of graphs and let $\mathcal{G}$ be a monotone and somewhere dense class of graphs. 
\begin{enumerate}
    \item If $\tw(\mathcal{H})< \infty$ then $\#\homsprob(\mathcal{H} \to \mathcal{G}) \in \mathsf{P}$. Moreover, if a tree decomposition of $H$ of width $t$ is given, then $\#\homsprob(\mathcal{H} \to \mathcal{G})$ can be solved in time $|H|^{O(1)}\cdot |V(G)|^{t+1}$. 
    \item If $\tw(\mathcal{H})=\infty$, then $\#\homsprob(\mathcal{H} \to \mathcal{G})$ is $\#\W[1]$-hard and, assuming ETH, cannot be solved in time $f(|H|)\cdot |G|^{o(\mathsf{tw}(H))}$ for any function $f$.
\end{enumerate}
\end{theorem}
\begin{proof}
The tractability result is well known~\cite{DiazST02,DalmauJ04}, so we only need to prove the hardness part. Recall that $\boxplus_k$ denotes the $k$-by-$k$ grid; see Figure~\ref{fig:wall_and_grid} for a depiction of $\boxplus_4$. Let $\boxplus := \{\boxplus_k~|~k\in \N\}$.
It is known that $\#\cphomsprob(\boxplus\to \all)$ is $\#\W[1]$-hard and, unless ETH fails, cannot be solved in time $f(k)\cdot |G|^{o(k)}$ for any function $f$ (see~\cite[Lemma 1.13 and 5.7]{Curticapean15} or \cite[Lemma 2.45]{Roth19}). As $\tw(\boxplus_k)=k$, the lower bound above can be written as $f(k)\cdot |G|^{o(\mathsf{tw}(\boxplus_k))}$.

Let $(\boxplus_k,(G,c))$ be the input to $\#\cphomsprob(\boxplus\to \all)$. Since $\mathcal{G}$ is somewhere dense and monotone, by Lemma~\ref{lem:nd_subdivisions} there is $r \in \No$ such that $\mathcal{G}$ contains the $r$-subdivision of every graph and thus, in particular, $G^r$.
Moreover, since $\tw(\mathcal{H})=\infty$ and $\mathcal{H}$ is minor-closed, by the Excluded-Grid Theorem~\cite{RobertsonS86-ExGrid} $\mathcal{H}$ contains every planar graph and thus in particular $\boxplus_k^r$. Clearly, $\boxplus_k^r$, $G^r$ and $c^r$ can be computed in polynomial time.
Moreover, by Lemma~\ref{lem:aa},
\[\#\homs{(\boxplus_k,\id_{\boxplus_k})}{(G,c)}=\#\homs{(\boxplus_k^r,\id_{\boxplus_k^r})}{(G^r,c^r)}\,.\]
Hence $\#\cphomsprob(\boxplus\to \all) \fptred \#\cphomsprob(\mathcal{H} \to \mathcal{G})$.
Since $\#\cphomsprob(\mathcal{H} \to \mathcal{G}) \fptred \#\homsprob(\mathcal{H} \to \mathcal{G})$ by Lemma~\ref{lem:hom_colreduction}, we conclude that $\#\homsprob(\mathcal{H} \to \mathcal{G})$ is $\#\W[1]$-hard.
For the conditional lower bound, observe that both reductions used above preserve the treewidth of the pattern (the first because treewidth is invariant under edge subdivision,\footnote{For example,
this invariance is in  Exercises 7.7 and 7.13 in~\cite{CyganFKLMPPS15}.} the second by Lemma~\ref{lem:hom_colreduction}). 
\end{proof}

\subsection{Monotone Pattern Classes}
The strengthening of  Theorem~\ref{thm:homs_warmup} to monotone pattern classes can be done by reduction from counting homomorphisms from a class of well-known graphs called \emph{walls}. 

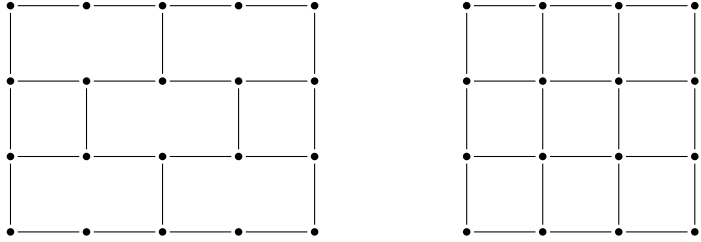
\begin{figure}[t]
        \centering
        \begin{tikzpicture}[scale=1]
            \centering
            
                \node[vertex] (11) at (1, 1) {};
                \node[vertex] (12) at (1, 2) {};
                \node[vertex] (13) at (1, 3) {};
                \node[vertex] (14) at (1, 4) {};
                
                \node[vertex] (21) at (2, 1) {};
                \node[vertex] (22) at (2, 2) {};
                \node[vertex] (23) at (2, 3) {};
                \node[vertex] (24) at (2, 4) {};
                
                \node[vertex] (31) at (3, 1) {};
                \node[vertex] (32) at (3, 2) {};
                \node[vertex] (33) at (3, 3) {};
                \node[vertex] (34) at (3, 4) {};
                
                \node[vertex] (41) at (4, 1) {};
                \node[vertex] (42) at (4, 2) {};
                \node[vertex] (43) at (4, 3) {};
                \node[vertex] (44) at (4, 4) {};
                
                \node[vertex] (51) at (5, 1) {};
                \node[vertex] (52) at (5, 2) {};
                \node[vertex] (53) at (5, 3) {};
                \node[vertex] (54) at (5, 4) {};

                \draw (11) -- (12); \draw (12) -- (13); \draw (13) -- (14);  
                \draw (51) -- (52); \draw (52) -- (53); \draw (53) -- (54);  
                
                \draw (11) -- (21); \draw (21) -- (31); \draw (31) -- (41); \draw (41) -- (51); 
                \draw (12) -- (22); \draw (22) -- (32); \draw (32) -- (42); \draw (42) -- (52); 
                \draw (13) -- (23); \draw (23) -- (33); \draw (33) -- (43); \draw (43) -- (53); 
                \draw (14) -- (24); \draw (24) -- (34); \draw (34) -- (44); \draw (44) -- (54); 
                
                \draw (31) -- (32); \draw (33) -- (34);  \draw (22) -- (23); \draw (42) -- (43); 
                
              \begin{scope}[xshift = +6cm]
              
                \node[vertex] (11) at (1, 1) {};
                \node[vertex] (12) at (1, 2) {};
                \node[vertex] (13) at (1, 3) {};
                \node[vertex] (14) at (1, 4) {};
                
                \node[vertex] (21) at (2, 1) {};
                \node[vertex] (22) at (2, 2) {};
                \node[vertex] (23) at (2, 3) {};
                \node[vertex] (24) at (2, 4) {};
                
                \node[vertex] (31) at (3, 1) {};
                \node[vertex] (32) at (3, 2) {};
                \node[vertex] (33) at (3, 3) {};
                \node[vertex] (34) at (3, 4) {};
                
                \node[vertex] (41) at (4, 1) {};
                \node[vertex] (42) at (4, 2) {};
                \node[vertex] (43) at (4, 3) {};
                \node[vertex] (44) at (4, 4) {};
                
                \draw (11) -- (12); \draw (12) -- (13); \draw (13) -- (14);
                \draw (21) -- (22); \draw (22) -- (23); \draw (23) -- (24);
                \draw (31) -- (32); \draw (32) -- (33); \draw (33) -- (34);
                \draw (41) -- (42); \draw (42) -- (43); \draw (43) -- (44);
                
                \draw (11) -- (21); \draw (21) -- (31); \draw (31) -- (41); 
                \draw (12) -- (22); \draw (22) -- (32); \draw (32) -- (42); 
                \draw (13) -- (23); \draw (23) -- (33); \draw (33) -- (43); 
                \draw (14) -- (24); \draw (24) -- (34); \draw (34) -- (44);

              \end{scope}
         \end{tikzpicture}
        \caption{The wall $W_{4,5}$ (\emph{left}) and the grid $\boxplus_4$ (\emph{right}).} 
        \label{fig:wall_and_grid}
    \end{figure}

\begin{definition}[Walls]
Let $k,\ell \in \N$. The \emph{wall} of height $k$ and length $\ell$, denoted by $W_{k,\ell}$, is the graph whose vertex set is $\{v_{i,j}: 1\leq i \leq k, 1\leq j\leq \ell\}$ and whose edge set contains:
\begin{itemize}\itemsep0pt
    \item $\{v_{i,j},v_{i,j+1}\}$ for all $1\leq i \leq k$ and $1\leq j\leq \ell-1$,
    \item $\{v_{i,1},v_{i+1,1}\}$ and $\{v_{i,\ell},v_{i+1,\ell}\}$ for all $1\leq i \leq k-1$
    \item $\{v_{i,j},v_{i+1,j}\}$ for all $1\leq i \leq k-1$ and $1\leq j\leq \ell$ such that $i+j$ is even.
\end{itemize}
\end{definition}
\noindent Figure~\ref{fig:wall_and_grid} depicts $W_{4,5}$ as an example. We let $\mathcal{W}:=\{W_{k,k}~|~k\in\N\}$ be the class of all walls.

The following structural property of large walls is due to Thomassen.\footnote{Note that walls are called grids in~\cite{Thomassen88}.}
\begin{lemma}[Proposition~3.2 in~\cite{Thomassen88}]\label{lem:erdos_posa_walls}
For every $k,r \in \N$, there exists $h(k,r) \in \N$ such that every subdivision of $W_{h(k,r),h(k,r)}$ contains as a subgraph a subdivision of $W_{k,k}$ in which each edge is subdivided a (positive) multiple of $r$ times.
\end{lemma}

The final ingredient of our proof for the classification of monotone pattern classes is given by Lemma~\ref{lem:nd_subdivision_strong}, which is an immediate consequence of Lemma~\ref{lem:nd_subdivisions}.
\begin{lemma}\label{lem:nd_subdivision_strong}
Let $\mathcal{G}$ be a monotone and somewhere dense class of graphs. There exists $r\in \No$ such that the following holds. Let $G$ be any graph and let $G'$ be any graph obtained from $G$ by subdividing each edge a (positive) multiple of $r$ times. Then $G'$ is contained in $\mathcal{G}$.
\end{lemma}
\begin{proof}
We show that the claim holds for the $r\in\No$ given by Lemma~\ref{lem:nd_subdivisions}. 
For this $r$, Lemma~\ref{lem:nd_subdivisions}
guarantees that for every graph~$H$, $H^r\in \mathcal{G}$.
Now let $G$ be any graph and label its edges $e_1,\dots,e_m$. Let $G'$ be any graph obtained from $G$ by subdividing each edge a (positive) multiple of $r$ times. Then there exist $d_1,\dots,d_m \in \N$ such that, for each $i\in[m]$, the edge $e_i$ is subdivided $d_ir$ times. Now let $\hat{G}$ be the graph obtained from $G$ by subdividing, for each $i\in[m]$, the edge $e_i$ just $d_i$ times.
It is immediate that $G'=\hat{G}^r$. Thus, by Lemma~\ref{lem:nd_subdivisions} and our choice of $r$, we have that $G'\in\mathcal{G}$.
\end{proof}

We are now ready to establish the main result of this section.
\begin{theorem}\label{thm:main_homs}
Let $\mathcal{H}$ be a monotone class of graphs and let $\mathcal{G}$ be a monotone and somewhere dense class of graphs. 
\begin{enumerate}
    \item If $\tw(\mathcal{H})< \infty$ then $\#\homsprob(\mathcal{H} \to \mathcal{G}) \in \mathsf{P}$. Moreover, if a tree decomposition of $H$ of width $t$ is given, then $\#\homsprob(\mathcal{H} \to \mathcal{G})$ can be solved in time $|H|^{O(1)}\cdot |V(G)|^{t+1}$. 
    \item If $\tw(\mathcal{H})=\infty$, then $\#\homsprob(\mathcal{H} \to \mathcal{G})$ is $\#\W[1]$-hard and, assuming ETH, cannot be solved in time $f(|H|)\cdot |G|^{o(\mathsf{tw}(H))}$ for any function $f$.
\end{enumerate}
\end{theorem}
\begin{proof}
The tractability result is well known~\cite{DiazST02,DalmauJ04}, so we only need to prove point 2. To this end, we will reduce from $\#\cphomsprob(\mathcal{W}\to \all)$. Walls clearly have grid minors of linear size, that is, there is a function $h\in\Theta(k)$ such that $W_{k,k}$ contains $\boxplus_{h(k)}$ as a minor. Furthermore, it is well-known that $\#\cphomsprob$ is minor-monotone (see e.g.\ \cite[Lemma 5.8]{Curticapean15} or \cite[Lemma 2.47]{Roth19}), hence $\#\cphomsprob(\boxplus\to \all)\fptred \#\cphomsprob(\mathcal{W}\to \all)$. Moreover the reduction is tight, in the sense that the lower bound for $\#\cphomsprob(\boxplus\to \all)$ shown in the proof of Theorem~\ref{thm:homs_warmup} transfers to $\#\cphomsprob(\mathcal{W}\to \all)$; hence $\#\cphomsprob(\mathcal{W}\to \all)$ is $\#\W[1]$-hard and, assuming ETH, it cannot be solved in time $f(k)\cdot |G|^{o(\mathsf{tw}(W_{k,k}))}$ for any function $f$.

Let us now construct the reduction $\#\cphomsprob(\mathcal{W}\to \all) \fptred \#\cphomsprob(\mathcal{H}\to\mathcal{G})$.
Let $r\in \No$ as given by Lemma~\ref{lem:nd_subdivision_strong}. We use the fact that $\tw(\mathcal{H})=\infty$ implies that $\mathcal H$ contains as minors all planar graphs; that is, for every planar graph $F$ there is a graph $H\in\mathcal{H}$ such that $F$ is a minor of $H$~\cite{RobertsonS86-ExGrid}. 
In particular, $\mathcal{H}$ contains all walls $W_{k,k}$ as minors. 
A graph $J$ is said to be a ``topological minor'' of a graph $H$ if there is a subdivision of $J$ that is isomorphic to a subgraph of $H$. 
Since walls have degree at most $3$, 
the fact that $\mathcal{H}$ contains all walls as minors implies that it also contains all walls as topological minors (see e.g. \cite[Proposition 1.7.3]{Diestel17}).  

Now let $W_{k,k}$ and $(G,c)$ be an input instance of $\#\cphomsprob(\mathcal{W}\to \all)$. Let $e_1,\ldots,e_\ell$ be the edges of $W_{k,k}$ in arbitrary order.  By Lemma~\ref{lem:erdos_posa_walls},   every subdivision of $W_{h(k,r),h(k,r)}$ contains as a subgraph a subdivision of $W_{k,k}$ in which each edge is subdivided a (positive) multiple of $r$ times. 
Since $\mathcal{H}$ contains $W_{k,k}$ as a topological minor,
there is a subdivision of $W_{k,k}$
that is isomorphic to a subgraph~$W'$ 
of a graph in $\mathcal{H}$. Since $\mathcal{H}$ is monotone, there are $W'\in \mathcal{H}$ and $d_1,\dots,d_\ell \in \No$ such that $W'$ is obtained from $W_{k,k}$ by subdividing $e_i$ precisely $d_ir$ times for each $i\in[\ell]$.   

We will now construct from $(G,c)$ a graph $G'$ and a surjective homomorphism~$c'$ from~$G'$ to~$W'$. 
For each edge $e=\{u,v\}$ of $G$ we proceed as follows. Since $c\in \homs{G}{W_{k,k}}$, then $\{c(u),c(v)\}=e_i$ for some $i\in[\ell]$. By the definition of $W'$,   $e_i$ was replaced by a path $c(u),x_1,\dots,x_{d_ir},c(v)$.
Hence, we replace the edge $e$ in $G$ by a path $u,w_1,w_2,\dots,w_{d_ir},v$, where the $w_j$ are fresh vertices. Furthermore, we extend the colouring $c$ to the colouring $c'$ by setting $c'(w_j):=x_j$ for each $j\in[d_ir]$. 
Since $c$ is surjective, so is $c'$. 
Also,

\[\#\homs{(W_{k,k},\id_{W_{k,k}})}{G,c} = \#\homs{(W',\id_{W'})}{(G',c')} \,.\]
By querying the oracle for $\#\cphomsprob(\mathcal{H}\to\mathcal{G})$ on the instance $((W',\id_{W'}),(G',c'))$ we can thus conclude our reduction.
This immediately implies $\#\W[1]$-hardness of $\#\cphomsprob(\mathcal{H}\to\mathcal{G})$. For the conditional lower bound, we observe that $W'$ has the same treewidth as $W_{k,k}$ since it is a subdivision of $W_{k,k}$, and that the size of $(G',c')$ is clearly bounded by $f(k)\cdot |G|^{O(1)}$ --- note that the $f$ depends on $\mathcal{H}$ which is, however, fixed.
A reduction to the uncoloured version via Lemma~\ref{lem:hom_colreduction} completes the proof.
\end{proof}

\noindent Theorem~\ref{thm:main_homs_intro} follows immediately from Theorem~\ref{thm:sparsity} and Theorem~\ref{thm:main_homs}. We conclude with a remark.

\begin{remark}
A strengthening of Theorem~\ref{thm:main_homs} to hereditary pattern classes $\mathcal{H}$ is not possible. Suppose for instance that $\mathcal{H}$ contains all complete graphs and $\mathcal{G}$ is the class of all bipartite graphs. Although $\mathcal{H}$ is hereditary and of unbounded treewidth, and $\mathcal{G}$ is monotone and somewhere dense, it is easy to see that $\#\homsprob(\mathcal{H} \to \mathcal{G})$ is trivial, since we can always output zero if $H\in\mathcal{H}$ has at least $3$ vertices.
When it comes to a sufficient and necessary condition for tractability in case of hereditary classes of patterns, we conjecture that \emph{induced grid minor size} might be the right candidate. However, even for very special cases, such as classes of degenerate host graphs (which are somewhere dense and monotone), it is still open whether induced grid minor size is the correct answer~\cite{BressanR21}. Thus, we leave the classification for hereditary classes of patterns as an open problem for further research. 
\end{remark}

\section{Counting Subgraphs}\label{sec:subgraphs}
This section is devoted to the proofs of Theorem~\ref{thm:intro_match}, Theorem~\ref{thm:subs_monotone_intro}, and Theorem~\ref{thm:subs_hereditary_intro}. We begin in Section~\ref{sec:subs_2} by analysing the problem of counting $k$-matchings in somewhere dense host graphs, and proving Theorem~\ref{thm:intro_match}; this is the most technical part. We then move on to prove Theorem~\ref{thm:subs_monotone_intro} and Theorem~\ref{thm:subs_hereditary_intro} in Section~\ref{sec:subs_3}.

\subsection{Counting Matchings: Proof of Theorem~\ref{thm:intro_match}}\label{sec:subs_2} 
A $k$-matching in a graph $G$ is a set $M\subseteq E(G)$ with $|M|=k$ and $e_1\cap e_2 = \emptyset$ for all $e_1\neq e_2$ in $M$. In other words, a $k$-matching in $G$ is a set of $k$ pairwise non-incident edges of $G$. Given a class of graphs $\mathcal{G}$, the problem $\#\match(\mathcal{G})$ asks, on input $k \in \N$ and a graph $G\in \mathcal{G}$, to compute the number of $k$-matchings in $G$; the parameter is $k$. We remark that $\#\match(\mathcal{G})=\#\subsprob(\mathcal{M} \to \mathcal{G})$ where $\mathcal{M}$ is the set of all $1$-regular graphs. The goal of this section is to prove that $\#\match(\mathcal{G})$ is hard whenever $\mathcal G$ is monotone and somewhere dense, i.e., the hardness part of Theorem~\ref{thm:intro_match}.

Before moving on, let us pin down some definitions and basic facts. Our analysis relies on the following ``coloured'' version of the graph tensor product, as in~\cite{PeyerimhoffRSSVW22}:
\begin{definition}\label{def:tensor}
Let $H$ be a graph, and let $(G_1,c_1)$ and $(G_2,c_2)$ be $H$-coloured graphs. The \emph{tensor product} $(G_1,c_1) \times (G_2,c_2)$ is the $H$-coloured graph $(\hat{G},\hat{c})$ defined by:
\begin{enumerate}\itemsep0pt
    \item[(T1)]  $V(\hat{G}) = \{(v_1,v_2) \in V(G_1) \times V(G_2)~|~c_1(v_1)=c_2(v_2) \}$.
    \item[(T2)]  $\{(u_1,u_2),(v_1,v_2)\}\in E(\hat{G})$ if and only if $\{u_1,v_1 \}\in E(G_1)$ and $\{u_2,v_2 \}\in E(G_2)$.
    \item[(T3)] $\hat{c}(v_1,v_2) = c_1(v_1)$ (equivalently by (T1), $\hat{c}(v_1,v_2)=c_2(v_2)$) for all $(v_1,v_2)\in V(\hat{G})$.
\end{enumerate}
\end{definition}
The crucial property of the tensor product is given by:\footnote{\label{fmark1}Proofs of Lemma~\ref{lem:tensor_linear_subgraphs} and Lemma~\ref{lem:MH_nonsingular} can also be found in Section 3.1 in an earlier version~\cite{RothSW21Arxiv} of~\cite{PeyerimhoffRSSVW22}.}

\begin{lemma}[\cite{PeyerimhoffRSSVW22}]\label{lem:tensor_linear_subgraphs}
If $H$ is a graph and $(F,c_F)$, $(G_1,c_1)$, $(G_2,c_2)$ are $H$-coloured graphs, then
\[\#\homs{(F,c_F)}{(G_1,c_1) \times (G_2,c_2)} = \#\homs{(F,c_F)}{(G_1,c_1)} \cdot \#\homs{(F,c_F)}{(G_2,c_2)}  \,.\]
\end{lemma}

The final ingredient we need is the non-singularity of a certain matrix whose entries count homomorphisms between fractured graphs. Formally, let $H$ be a graph. The square matrix $M_H$ has its rows and columns indexed by the fractures of $H$, and its entries satisfy:
\begin{equation}\label{eq:MHdef}
    M_H[\rho,\sigma]:= \#\homs{(\fracture{H}{\rho},c_\rho)}{(\fracture{H}{\sigma},c_\sigma)}\,,
\end{equation}
where $c_\rho$ and $c_\sigma$ are the canonical $H$-colourings of the fractured graphs $\fracture{H}{\rho}$ and $\fracture{H}{\sigma}$ (see Definition~\ref{def:canonical_col} and Observation~\ref{obs:canonical_col}). By ordering the columns and rows of $M_H$ along a certain lattice, the following property was established in previous work.\footnote{See Footnote~\ref{fmark1}.}

\begin{lemma}[\cite{PeyerimhoffRSSVW22}]\label{lem:MH_nonsingular}
For each graph $H$, the matrix $M_H$ is nonsingular.
\end{lemma}

If $\mathcal{G}$ is closed under \emph{uncoloured} tensor products\footnote{The adjacency matrix of the tensor product of two uncoloured graphs $G$ and $F$ is the Kronecker product of the adjacency matrices of $G$ and $F$.}, then the hardness result can be achieved by applying the reduction of~\cite{CurticapeanDM17} verbatim. However, that reduction fails if $\mathcal{G}$ is not closed under uncoloured tensor products, and this closure property is very restrictive. Consider for example the class $\mathcal{G}$ of square-free graphs, i.e., graphs that do not contain the $4$-cycle $C_4$ as a subgraph. Then $\mathcal G$ is clearly monotone and, since it contains the $3$-subdivision of every graph, it is also somewhere dense by Lemma~\ref{lem:nd_subdivisions}. However, $\mathcal G$ is not closed under (uncoloured) tensor products: the path on $2$ edges $P_2$ is in $\mathcal{G}$, but $P_2\times P_2 \notin \mathcal G$ since it contains a $C_4$.

The main insight of this section is a weakened closure property for monotone and somewhere dense graph classes, established in the lemma below. Combined with the characterisation of somewhere dense graph classes via $r$-subdivisions (Lemma~\ref{lem:nd_subdivisions}), this property implies that any monotone and somewhere dense class is closed under tensor products of subdivisions of \emph{coloured} graphs.

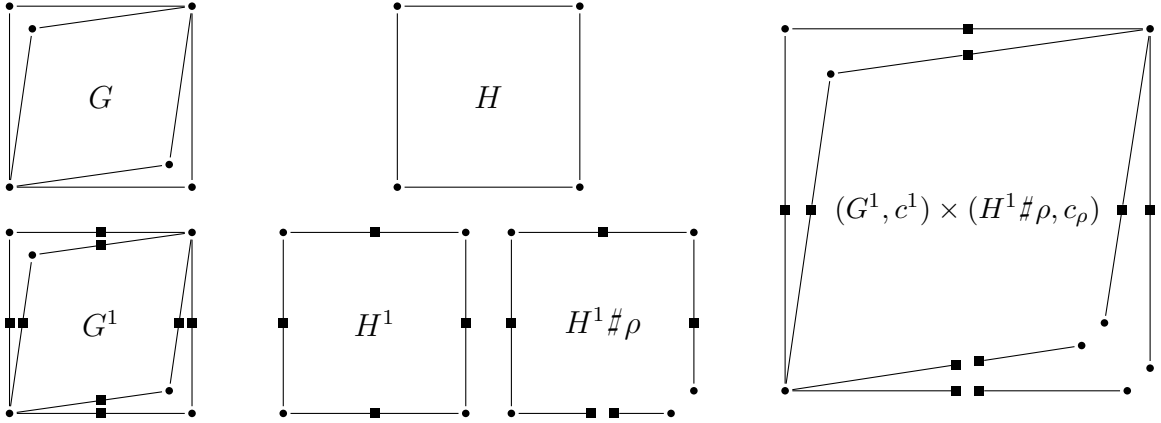
\begin{figure}[t]
        \centering
        \begin{tikzpicture}[scale=0.6]
            \centering
            \begin{scope}
            
                \node[vertex] (U1) at (0, -0) {};
                \node[vertex] (U2) at (0, -4) {};
                \node[vertex] (V1) at (-4, 0) {};
                \node[vertex] (V2) at (-4, -4) {};
                \node[vertex] (V11) at (-3.5, -0.5) {};
                \node[vertex] (U21) at (-0.5, -3.5) {};
               
                \draw (U1) -- (V1); \draw (U1) -- (U2); \draw (V2) -- (V1); \draw (V2) -- (U2); 
                \draw (U21) -- (V2); \draw (U21) -- (U1); \draw (V11) -- (V2); \draw (V11) -- (U1);
                
                \node (e1) at (-2.7, .1) {};
                \node (e2) at (0.7, -2.1) {};

                \node at (-2,-2) {\large $G$};

                \node[vertex] (U1) at (8.5, 0) {};
                \node[vertex] (U2) at (8.5, -4) {};
                \node[vertex] (V1) at (4.5, 0) {};
                \node[vertex] (V2) at (4.5, -4) {};
               \draw (U1) -- (V1); \draw (U1) -- (U2); \draw (V2) -- (V1); \draw (V2) -- (U2);

                \node (e1) at (3.3, .1) {};
                \node (e2) at (6.7, -2.1) {};
               
                \node at (6.5,-2) {\large $H$};
               
            \end{scope}
            
            \begin{scope}[yshift =-5cm]
            
                \node[vertex] (U1) at (0, -0) {};
                \node[vertex] (U2) at (0, -4) {};
                \node[vertex] (V1) at (-4, 0) {};
                \node[vertex] (V2) at (-4, -4) {};
                \node[vertex] (V11) at (-3.5, -0.5) {};
                \node[vertex] (U21) at (-0.5, -3.5) {};
                
                \node[inner sep =1.75pt, fill] (M) at (-2, 0) {};
                \node[inner sep =1.75pt, fill] (M) at (-2, -2/7) {};
                
                \node[inner sep =1.75pt, fill] (M) at (-2, -4) {};
                \node[inner sep =1.75pt, fill] (M) at (-2, -4 +2/7) {};
                
                \node[inner sep =1.75pt, fill] (M) at (0, -2) {};
                \node[inner sep =1.75pt, fill] (M) at (-2/7, -2) {};
                
                \node[inner sep =1.75pt, fill] (M) at (-4, -2) {};
                \node[inner sep =1.75pt, fill] (M) at (-4+2/7, -2) {};
               
                \draw (U1) -- (V1); \draw (U1) -- (U2); \draw (V2) -- (V1); \draw (V2) -- (U2); 
                \draw (U21) -- (V2); \draw (U21) -- (U1); \draw (V11) -- (V2); \draw (V11) -- (U1);
                
                \node (e1) at (-2.7, .1) {};
                \node (e2) at (0.7, -2.1) {};

                \node at (-2,-2) {\large $G^1$};

                \node[vertex] (U1) at (6, 0) {};
                \node[vertex] (U2) at (6, -4) {};
                \node[vertex] (V1) at (2, 0) {};
                \node[vertex] (V2) at (2, -4) {};
                \draw (U1) -- (V1); \draw (U1) -- (U2); \draw (V2) -- (V1); \draw (V2) -- (U2);
                \node[inner sep =1.75pt, fill] (M) at (4, 0) {};
                
                \node[inner sep =1.75pt, fill] (M) at (4, -4) {};
                
                \node[inner sep =1.75pt, fill] (M) at (6, -2) {};
                
                \node[inner sep =1.75pt, fill] (M) at (2, -2) {};

                \node at (4,-2) {\large $H^1$};

                \node[vertex] (U1) at (11, 0) {};
                \node[vertex] (U2) at (11, -3.5) {};
                \node[vertex] (U22) at (10.5, -4) {};
                \node[vertex] (V1) at (7, 0) {};
                \node[vertex] (V2) at (7, -4) {};
                
                \node[inner sep =1.75pt, fill] (M1) at (9, 0) {};
                
                \node[inner sep =1.75pt, fill] (M2) at (9.25, -4) {};
                \node[inner sep =1.75pt, fill] (M12) at (8.75, -4) {};
                
                \node[inner sep =1.75pt, fill] (M3) at (11, -2) {};
                
                \node[inner sep =1.75pt, fill] (M4) at (7, -2) {};

               \draw (U1) -- (V1); \draw (U1) -- (U2); \draw (V2) -- (V1); \draw (V2) -- (M12); \draw (M2) -- (U22);
               
               \node at (9,-2) {\large $\fracture{H^1}{\rho}$};

            \end{scope}
           
            \begin{scope}[xshift =11cm, yshift=-0.5cm]
              \node[vertex] (U1) at (10, 0) {};
                \node[vertex] (U2) at (10, -7.5) {};
                \node[vertex] (U22) at (9.5, -8) {};
                \node[vertex] (W2) at (9, -6.5) {};
                \node[vertex] (W22) at (8.5, -7) {};
                \node[vertex] (V1) at (2, 0) {};
                \node[vertex] (W1) at (3, -1) {};
                \node[vertex] (V2) at (2, -8) {};
                
                \node[inner sep =1.75pt, fill] (M1) at (6, 0) {};
                \node[inner sep =1.75pt, fill] (M1N1) at (6, -4/7) {};
                
                \node[inner sep =1.75pt, fill] (M2) at (6.25, -8) {};
                \node[inner sep =1.75pt, fill] (M2N1) at (6.25, -191/26) {};
                \node[inner sep =1.75pt, fill] (M2N2) at (5.75, -193/26) {};
                
                \node[inner sep =1.75pt, fill] (M12) at (5.75, -8) {};
                
                \node[inner sep =1.75pt, fill] (M3) at (10, -4) {};
                \node[inner sep =1.75pt, fill] (M3N1) at (61/6.5, -4) {};
                
                \node[inner sep =1.75pt, fill] (M4) at (2, -4) {};
                \node[inner sep =1.75pt, fill] (M4N1) at (2+4/7, -4) {};

                \node (e1) at (3.3, .1) {};
                \node (e2) at (6.7, -2.1) {};
       
               \draw (U1) -- (V1); \draw (U1) -- (U2); \draw (V2) -- (V1); \draw (V2) -- (M12); \draw (M2) -- (U22);
               
               \draw (M3N1) -- (W2); \draw (M3N1) -- (U1); \draw (M1N1) -- (U1); \draw (M1N1) -- (W1); \draw (M4N1) -- (W1);
               \draw (M4N1) -- (V2); \draw (M2N2) -- (V2); \draw (M2N1) -- (W22);
               
               \node at (6,-4) { $(G^1,c^1) \times (\fracture{H^1}{\rho},c_\rho)$};
            \end{scope}
         \end{tikzpicture}
        \caption{the tensor product of the $H^1$-coloured graphs $(G^1,c^1)$ and  $(\fracture{H^1}{\rho},c_\rho)$.} 
        \label{fig:main_tensor_subs}
    \end{figure}

\begin{lemma}\label{lem:main_tensor_subs}
Let $r \in \No$, let $H$ be a graph without isolated vertices, and let $(G,c)$ be an $H$-coloured graph on $n$ vertices, and let $\rho$ be a fracture of~$H^r$. Then
$(G^r,c^r) \times (\fracture{H^r}{\rho},c_\rho)$
is a subgraph of the $r$-subdivision of a complete graph of order $O(k n)$, where $k=|E(H)|$ and the constants in the $O()$ notation depend only on $r$.
\end{lemma}
\begin{proof}
Let $T = (G^r,c^r) \times (\fracture{H^r}{\rho},c_\rho)$;
see Figure~\ref{fig:main_tensor_subs} for an example. The claim follows from Claims~1, 2, and 3 below, with Claim 3 applied to $F=T$.

\vspace*{5pt}
\noindent{\bf Claim 1.} $|V(T)| = O(k n)$. Straightforward since $G^r$ is a subgraph of $K_n^r$. 
 
\vspace*{5pt}
\noindent{\bf Claim 2:} if $x$ and $y$ are distinct vertices of~$T$ of degree at least $3$, then the length of any simple path from $x$ to $y$ is a multiple of $r+1$.

To prove this, recall that $T$ is $H^r$-coloured by $\hat{c}$ from Definition~\ref{def:tensor}, and that $V(H^r)$ can be partitioned into $V(H)$ and a set $S$ of $k r$ fresh subdivision vertices. Let $(u,v)$ be a vertex of $T$ such that $\hat{c}(u,v) = s\notin V(H)$, that is, $(u,v)$ is coloured with a subdivision vertex $s$. We show that $(u,v)$ has degree at most $2$ in~$T$. Let $s_1$ and $s_2$ be the two neighbours of $s$ in $H^r$. By the construction of $(G^r,c^r)$, $u$ has exactly two neighbours in~$G^r$, say $u_1$ and $u_2$. Furthermore, $c^r(u_1)= s_1$ and $c^r(u_2)=s_2$.  Since $s$ has degree $2$ in~$H^r$, there are only two cases for $\rho_s$.
\begin{itemize}\itemsep0pt
\item  Case 1: $\rho_s=\{B\}$ where $B=\{\{s,s_1\},\{s,s_2\}\}$. In this case $s^B$ is the only vertex of $\fracture{H}{\rho}$ that is coloured by $c_\rho$ with $s$. Since $\hat{c}(u,v) = s$ implies $c_\rho(v)=s$, we conclude that $v=s^B$. Hence $(u,v)$ has exactly two neighbours in~$T$,  $(u_1,s_1^{B_1})$ and $(u_2,s_2^{B_2})$, where $B_1$ and $B_2$ are the blocks of $\rho_{s_1}$ and $\rho_{s_2}$ containing respectively $\{s,s_1\}$ and $\{s,s_2\}$.
\item Case 2: $\rho_s=\{B,B'\}$ where $B=\{\{s,s_1\}\}$ and $B'=\{\{s,s_2\}\}$.
In this case $s^B$ and $s^{B'}$ are the only two vertices of $\fracture{H}{\rho}$ that are coloured by $c_\rho$ with $s$. Since $\hat{c}(u,v) = s$ implies $c_\rho(v)=s$, we conclude that  $v\in \{s^B,s^{B'}\}$. Assume that $v=S^B$; the other case is symmetric. Then the only neighbour of~$(u,v)$ in~$T$ is   $(u_1,s_1^{B_1})$, where $B_1$ is the block of $\rho_{s_1}$ that contains the edge $\{s,s_1\}$.
\end{itemize}
We conclude that the only vertices $(u,v)$ of degree at least $3$ in $T$ satisfy  $\hat{c}(u,v) \in V(H)$, implying that $c^r(u)\in V(H)$ and thus, by the definition of $c^r$, that $u\in V(G)$, hence $u$ is not a subdivision vertex. 
The claim follows since the length of every simple path between two non-subdivision vertices $u_1$ and $u_2$ in $G^r$ is a multiple of $(r+1)$, and since $T$ can be obtained from $(G^r,c^r)$ by splitting vertices.

{\bf Claim 3:} if $F$ is a graph where the length of any simple path between two vertices of degree at least $3$ is a multiple of $(r+1)$, then $F$ is a subgraph of the $r$-subdivision of a complete graph of order $O(|V(F)|)$.

Note first that we can deal with each connected component of $F$ separately. Furthermore, the claim is clearly true if $F$ is just a path (of any length). For what follows we can hence assume that $F$ is connected and not isomorphic to a path.
We say that a path~$P$ in~$F$ is \emph{extendable} if its internal vertices have degree~$2$, one endpoint $s_P$
(the ``startpoint'') has degree~$1$, and the other endpoint has degree at least~$3$. 
If $P$ has length $\ell_P$, 
then its \emph{extension length} is the smallest $\ell'_P \in \No$ such that $\ell_P+\ell'_P$ is a multiple of~$r+1$.
Let $F'$ be the graph formed from~$F$ by considering 
every extendable path~$P$ and adding a new length-$\ell'_P$ path from $s_P$ (adding $\ell'_P$ fresh vertices to make up this path). Observe that, for every pair of non-isolated vertices $u'$ and $v'$ of $F'$, if both $u'$ and $v'$ have degree not equal to $2$, then the length of every simple path from $u'$ to $v'$ in $F'$ is a multiple of $(r+1)$.
Therefore $F'$ is a subgraph of the $r$-subdivision of a complete graph of order at most $O(|V(F')|) =  O(|V(F)|)$, where the constants depend only on $r$. Moreover $F$ is by construction a subgraph of~$F'$, which this concludes the proof of the claim.
\end{proof}

To establish the hardness of $\#\match(\mathcal{G})$, we first consider an edge-coloured version. Let $G$ be a graph and $k \in \N$. A $k$-coloring of $E(G)$ is a map $c:E(G)\to \{1,\ldots,k\}$. A matching $M \subseteq E(G)$ is \emph{edge-colorful} under if for every colour in $\{1,\ldots,k\}$ there is precisely one element of $M$ with that colour.
\begin{definition}[$\#\colmatch(\mathcal{G})$]
Let $\mathcal{G}$ be a class of graphs. The problem $\#\colmatch(\mathcal{G})$ asks, on input $k \in \N$, a graph $G\in\mathcal{G}$, and a $k$-coloring $c$ of $E(G)$, to compute the number of edge-colorful $k$-matchings in $G$. The problem is parameterised by $k$.
\end{definition}

\begin{lemma}\label{lem:main_colmatch}
Let $\mathcal{G}$ be a monotone somewhere dense class of graphs. Then $\#\colmatch(\mathcal{G})$ is $\#\mathsf{W}[1]$-hard and, assuming ETH, cannot be solved in time $f(k)\cdot |G|^{o(k/\log k)}$ for any function $f$.
\end{lemma}
\begin{proof}
Let $\mathcal{H}$ be a class of $3$-regular expander graphs. Both the treewidth and the number of edges of the elements of $\mathcal H$ grow linearly in the number of vertices; that is, $|E(H)|\in\Theta(|V(H)|)$ and $\tw(H)\in \Theta(|V(H)|)$ for all $H\in\mathcal{H}$~(see, e.g., \cite{GroheM09}). Hence theorems~\ref{thm:cphomshard} and~\ref{thm:cybt} imply that $\#\cphomsprob(\mathcal{H}\to \all)$ is $\#\mathsf{W}[1]$-hard and, assuming ETH, cannot be solved in time $f(|H|)\cdot |G|^{o(|H|/\log |H|)}$ for any function $f$. We will now show that $\#\cphomsprob(\mathcal{H}\to \all) \fptred \#\colmatch(\mathcal{G})$.
    
Let $H\in\mathcal{H}$ and $(G,c)$ be the input of $\#\cphomsprob(\mathcal{H}\to \all)$. By Lemma~\ref{lem:nd_subdivisions}, there is $r\in\No$ such that $G^r\in\mathcal{G}$ for all $G \in \all$. Construct then $H^r$ and $(G^r,c^r)$, which clearly takes polynomial time. Let $k=|E(H^r)|$; clearly $k\in O(|H|)$ where the constants depend only on~$r$. Now, by Lemma~\ref{lem:hom_subdiv},
    \[\#\homs{(H,\id_H)}{(G,c)}=\#\homs{(H^r,\id_{H^r})}{(G^r,c^r)}\,.\]
Next, we view surjectively $H^r$-coloured graphs $(\Tilde{G},\Tilde{c})$ also as edge-coloured graphs where every edge $e=\{u,v\}$ is mapped to the colour $\{\Tilde{c}(u),\Tilde{c}(v)\}$. This allows us to invoke the results of~\cite{PeyerimhoffRSSVW22} and deduce what follows.\footnote{In~\cite{PeyerimhoffRSSVW22}, the number of edge-colourful $k$-matchings of $G$ is denoted by $\#\coledgesubs{\Phi,k}{G}$, where $\Phi$ is the graph property of being a matching. The identities (\ref{eq:main_subs_1}) and (\ref{eq:main_subs_2}) are immediate consequences of Lemma~4.1 and Corollary~4.3 in~\cite{PeyerimhoffRSSVW22} (see also Lemma~3.1 and Corollary~3.3 in an earlier version~\cite{RothSW21Arxiv} of~\cite{PeyerimhoffRSSVW22}).}
    
First, there is a unique function $a$ from fractures of $H^r$ to rationals such that, for every surjectively $H^r$-coloured graph $(\Tilde{G},\Tilde{c})$, the number of edge-colourful $k$-matchings of $(\Tilde{G},\Tilde{c})$ is:
\begin{equation}\label{eq:main_subs_1}
    \sum_\rho a(\rho)\cdot \#\homs{(\fracture{H^r}{\rho},c_\rho)}{(\Tilde{G},\Tilde{c})}\,,
\end{equation}
where the sum is over all fractures of $H^r$. Additionally, $a$ satisfies:
\begin{equation}\label{eq:main_subs_2}
       a(\top) =  \prod_{v\in V(H^r)} (-1)^{\mathsf{deg}(v)-1} \cdot (\mathsf{deg}(v)-1)! \,,
\end{equation}
where $\top$ is the coarsest fracture, that is, for each $v\in V(H^r)$ the partition $\top_v$ only contains a singleton block (and therefore $\fracture{H^r}{\top}=H^r$). In particular, it is easy to see that \[a(\top)=\pm 2^{|V(H)|}\neq 0\,.\]

Now let $\sigma$ be a fracture of $H^r$. Considering~\eqref{eq:main_subs_1} with $(\tilde G, \tilde c) = (G^r,c^r) \times (\fracture{H^r}{\sigma},c_\sigma)$ and applying Lemma~\ref{lem:tensor_linear_subgraphs}, the number of colorful $k$-matchings in $(G^r,c^r) \times (\fracture{H^r}{\sigma},c_\sigma)$ equals:
\begin{align}\label{eq:colorful_sum_rho}
    \sum_\rho a(\rho)\cdot \#\homs{(\fracture{H^r}{\rho},c_\rho)}{(G^r,c^r)} \cdot \#\homs{(\fracture{H^r}{\rho},c_\rho)}{(\fracture{H^r}{\sigma},c_\sigma)}
\end{align}
By Lemma~\ref{lem:main_tensor_subs}, $(G^r,c^r) \times (\fracture{H^r}{\sigma},c_\sigma)$ is a subgraph of the $r$-subdivision of a complete graph, which is in $\mathcal G$ by our choice of $r$. Since $\mathcal G$ is monotone this implies $(G^r,c^r) \times (\fracture{H^r}{\sigma},c_\sigma) \in \mathcal G$, too. Hence, if we have an oracle for $\#\colmatch(\mathcal{G})$, then we can compute the value of~\eqref{eq:colorful_sum_rho}, while $\#\homs{(\fracture{H^r}{\rho},c_\rho)}{(\fracture{H^r}{\sigma},c_\sigma)}$ can obviously be computed in a time that is a function of $|H|$ and $r$.
Thus, by letting $\mathsf{coeff}(\rho):=a(\rho)\cdot \#\homs{(\fracture{H^r}{\rho},c_\rho)}{(G^r,c^r)}$, in FPT time we obtain a system of linear equations with unknowns $\mathsf{coeff}(\rho)$ and whose matrix is $M_{H^r}$, see~\eqref{eq:MHdef}. By Lemma~\ref{lem:MH_nonsingular} $M_{H^r}$ is nonsingular, hence by solving the system we can retrieve:
    \[\mathsf{coeff}(\top) =a(\top)\cdot \#\homs{(\fracture{H^r}{\top},c_\top)}{(G^r,c^r)} = a(\top)\cdot \#\homs{(H^r,\mathsf{id}_{H^r})}{(G^r,c^r)} \,.\]
Since $a(\top)\neq 0$, we can divide by $a(\top)$ and recover $\#\homs{(H^r,\mathsf{id}_{H^r})}{(G^r,c^r)}$ as desired. This concludes the parameterized reduction to $\#\colmatch(\mathcal{G})$ and proves the thesis.
\end{proof}

With the hardness results for $\#\colmatch(\mathcal{G})$ above, we can finally obtain our complexity dichotomy for $\#\match(\mathcal{G})$. First, we prove:
\begin{theorem}\label{thm:main_matchings}
Let $\mathcal{G}$ be a monotone somewhere dense class of graphs. Then $\#\match(\mathcal{G})$ is $\#\mathsf{W}[1]$-hard and, assuming ETH, cannot be solved in time $f(k)\cdot |G|^{o(k/\log k)}$ for any function $f$.
\end{theorem}
\begin{proof}
A well-known application of inclusion-exclusion (see, e.g., \cite[Lemma~1.34]{Curticapean15}) yields a parameterized reduction from $\#\colmatch(\mathcal{G})$ to $\#\match(\mathcal{G}')$ that preserves the parameter, where $\mathcal{G}'$ is the class of all subgraphs of $\mathcal G$. By monotonicity $\mathcal G' = \mathcal G$, so the claim of Lemma~\ref{lem:main_colmatch} holds for $\#\match(\mathcal{G})$, too.
\end{proof}
\noindent Finally, we obtain:
\begin{corollary}[Theorem~\ref{thm:intro_match}, restated]\label{xxx}
Let $\mathcal{G}$ be a monotone class of graphs and assume that ETH holds. Then $\#\match(\mathcal{G})$ is fixed-parameter tractable if and only if $\mathcal{G}$ is nowhere dense. In particular, if $\mathcal{G}$ is nowhere dense then $\#\match(\mathcal{G})$ can be solved in time $f(k)\cdot|V(G)|^{1+o(1)}$ for some computable function~$f$; otherwise $\#\match(\mathcal{G})$ cannot be solved in time $f(k)\cdot |G|^{o(k/\log k)}$ for any function $f$.
\end{corollary}
\begin{proof}
Immediate from Theorem~\ref{thm:sparsity} and Theorem~\ref{thm:main_matchings}.
\end{proof}

\begin{remark}
Unless $\#\mathrm{P}=\mathrm{P}$, Corollary~\ref{xxx} / Theorem~\ref{thm:intro_match}  cannot be strengthened to achieve polynomial time tractability of $\#\match(\mathcal{G})$ for nowhere dense and monotone $\mathcal{G}$. Let indeed $\mathcal G$ be the class of all $K_8$-minor-free graphs. Then $\mathcal{G}$ is clearly monotone, and since it does not contain the subdivisions of cliques larger than $7$, it is also nowhere dense by Lemma~\ref{lem:nd_subdivisions}. However, as shown recently by Curticapean and Xia~\cite{CurticapeanX22}, counting perfect matchings (i.e., $k$-matchings with $k=n/2$) in $K_8$-minor-free graphs is $\#\mathrm{P}$-hard.
\end{remark}

\subsection{Counting Subgraphs: Proofs of Theorems~\ref{thm:subs_hereditary_intro} and~\ref{thm:subs_monotone_intro}} \label{sec:subs_3}
Equipped with our hardness results for counting $k$-matchings, we move towards proving hardness for counting subgraphs. 
\begin{theorem}[Theorem~\ref{thm:subs_hereditary_intro}, restated]
Let $\mathcal{H}$ and $\mathcal{G}$ be graph classes such that $\mathcal{H}$ is hereditary and $\mathcal{G}$ is monotone. Then Table~\ref{tab:resultsSubsHereditary} exhaustively classifies the complexity of $\#\subsprob(\mathcal{H} \to \mathcal{G})$.
\end{theorem}
\begin{proof}
Let us first show that the cases for $\mathcal H$ and $\mathcal G$ in Table~\ref{tab:resultsSubsHereditary} are  exhaustive and mutually exclusive. For $\mathcal G$ this is straightforward. For $\mathcal H$, the first row and the rest are mutually exclusive and exhaustive, since rows 2, 3 and 4 all imply $\m(\mathcal H)=\infty$. To see that rows 2, 3, and 4 are mutually exclusive and exhaustive for  $\m(\mathcal{H})=\infty$, note that in that case Theorem~\ref{thm:ramsey_match} implies that at least one of $\mind(\mathcal{H})$, $\bind(\mathcal{H})$ and $\omega(\mathcal{H})$ is unbounded.
\begin{table}[t]
    \begin{tabularx}{\textwidth}{l|cccc}
          \begin{tabular}{c}
             $~$
        \end{tabular}&\begin{tabular}{c}
             $\mathcal{G}$
             n.\ dense
        \end{tabular} &\begin{tabular}{c}
             $\mathcal{G}$
             s.\ dense\\
             $\omega(\mathcal{G})=\infty$
        \end{tabular}& \begin{tabular}{c}
             $\mathcal{G}$
             s.\ dense\\
             $\omega(\mathcal{G})< \infty$\\  
             ${\beta}(\mathcal{G})=\infty$
        \end{tabular}&\begin{tabular}{c}
             $\mathcal{G}$
             s.\ dense\\
             $\omega(\mathcal{G})< \infty$\\ 
             ${\beta}(\mathcal{G})< \infty$
        \end{tabular}\\
        & \\[\dimexpr-\normalbaselineskip+5pt]
        \hline
        & \\[\dimexpr-\normalbaselineskip+5pt]
        \begin{tabular}{l}
             $\m(\mathcal{H})< \infty$
        \end{tabular} & P & P & P & P \\
        & \\[\dimexpr-\normalbaselineskip+5pt]
        \hline
        & \\[\dimexpr-\normalbaselineskip+5pt]
        \begin{tabular}{l}
             $\mind(\mathcal{H})= \infty$
        \end{tabular} & FPT & hard & hard & hard \\
        & \\[\dimexpr-\normalbaselineskip+5pt]
        \hline
        & \\[\dimexpr-\normalbaselineskip+5pt]
        \begin{tabular}{l}
            $\mind(\mathcal{H})< \infty$\\
            $\bind(\mathcal{H})= \infty$
        \end{tabular} & P & hard$^\dagger$ & hard$^\dagger$ & P \\
        & \\[\dimexpr-\normalbaselineskip+5pt]
        \hline
        & \\[\dimexpr-\normalbaselineskip+5pt]
        \begin{tabular}{l}
             $\mind(\mathcal{H}),\bind(\mathcal{H})< \infty$\\
            $\omega(\mathcal{H})=\infty$
        \end{tabular} & P & hard$^\dagger$ & P & P 
        \end{tabularx}
     \caption[caption]{\small The complexity of $\#\subsprob(\mathcal{H}\to \mathcal{G})$ for hereditary $\mathcal{H}$ and monotone $\mathcal{G}$ (Theorem~\ref{thm:subs_hereditary_intro}). P and FPT stand respectively for polynomial-time tractability and fixed-parameter tractability, hard means $\#\W[1]$-hard and without an algorithm running in time $f(|H|)\cdot |G|^{o(|V(H|)/ \log |V(H)|)}$ for any function $f$ unless ETH fails, and hard$^\dagger$ means the same but with a lower bound of $f(|H|)\cdot |G|^{o(|V(H)|)}$. The FPT entry cannot be strengthened to P unless $\mathrm{P}=\#\mathrm{P}$, see Remark~\ref{rem:realFPT}.
     }
      \label{tab:resultsSubsHereditary}
\end{table}

Let us now prove the entries of Table~\ref{tab:resultsSubsHereditary}.
The first row is due to Curticapean and Marx~\cite{CurticapeanM14}, and the FPT result in the first column follows from Theorem~\ref{thm:sparsity}. 
The intractability results in the second row follow from Theorem~\ref{thm:main_matchings} and the fact that $\mind(\mathcal{H})=\infty$ implies that $\mathcal{H}$ contains all matchings (since $\mathcal H$ is hereditary).
For the second column, note that $\omega(\mathcal{G})=\infty$ and $\mathcal{G}$ being monotone implies that $\mathcal{G}=\all$; the dichotomy of Curticapean and Marx~\cite{CurticapeanM14} then applies again.\footnote{\label{fn:2}The tight conditional lower bounds in the second column follow from the fact that the respective entries subsume counting $k$-cliques in arbitrary graphs, and counting $k$-by-$k$ bicliques in bipartite graphs. The tight bound of the former was shown in~\cite{Chenetal05,Chenetal06}, and the tight bound of the latter was implicitly shown in~\cite{CurticapeanM14}, and explicitly in~\cite{DorflerRSW22}; while~\cite{DorflerRSW22} studies \emph{induced} subgraphs in bipartite graphs, we note that all bicliques in a bipartite graph must be induced.}
Next, we prove the remaining entries.
 
\vspace*{5pt}
{\noindent $\bullet$ Row 3, Column 3:
if ${\beta(\mathcal{G})}=\bind(\mathcal{H})=\infty$ then $\#\subsprob(\mathcal{H}\to\mathcal{G})$ is hard.}
Since $\mathcal{H}$ is hereditary, it  contains all bicliques. Since $\mathcal{G}$ is monotone, it contains all bipartite graphs. Hence $\#\subsprob(\mathcal{H}\to\mathcal{G})$ is at least as hard as counting $k$-by-$k$ bicliques in bipartite graphs, which is known to be hard~\cite{CurticapeanM14}.\footnote{See Footnote~\ref{fn:2} for the tight conditional lower bound.}

\vspace*{5pt}
{\noindent $\bullet$ Row 3, Column 4: if $\mind(\mathcal{H}),\omega(\mathcal{G}),\beta(\mathcal{G}) < \infty$ then $\#\subsprob(\mathcal{H}\to\mathcal{G})$ is in polynomial time.} 
Let $(H,G)$ be the input of $\#\subsprob(\mathcal{H}\to\mathcal{G})$. If $\omega(H)>\omega(\mathcal G)$ or $\bind(H)>\beta(\mathcal G)$, then we can output $0$. We can thus restrict the problem to those $H$ such that $\omega(H)\le \omega(\mathcal G)$ and $\bind(H)\le \beta(\mathcal G)$. Recall that $\mind(H) \le \mind(\scH) < \infty$.
By the contrapositive of Theorem~\ref{thm:ramsey_match}, there is a monotonically increasing function $R$ such that: 
    \[\m(H) \leq R(\mind(H),\omega(H),\bind(H)) \leq R(\mind(\scH),\omega(\scG),\beta(\scG)) < \infty,\]
where the second inequality holds by monotonicity of $R$ and the third one by the boundedness of all three arguments. We therefore obtain polynomial time as in the first row.
 
\vspace*{5pt}
{\noindent $\bullet$ Row 4, Columns 3 and 4:
if $\mind(\mathcal{H}),\bind(\mathcal{H}),\omega(\mathcal{G}) < \infty$, then $\#\subsprob(\mathcal{H}\to\mathcal{G})$ is in polynomial time. }

Let $(H,G)$ be the input of $\#\subsprob(\mathcal{H}\to\mathcal{G})$. If $\omega(H)>\omega(\mathcal G)$ then we output $0$, hence we can assume that $\omega(H)\leq \omega(\mathcal G)$. Similarly to the previous case, we then obtain polynomial time since
\[\m(H) \leq R(\mind(H),\omega(H),\bind(H)) \leq R(\mind(H),\omega(\mathcal G),\bind(H)) < \infty.\]

\vspace*{5pt}
{\noindent $\bullet$ Rows 3 and 4, Column 1: $\#\subsprob(\mathcal{H}\to\mathcal{G})$ is in polynomial time.} 
We show that 
$\omega(\mathcal{G}),\beta(\mathcal{G})< \infty$; then the same arguments used for Rows~3 and~4 of Column~4 apply.
Suppose by contradiction that $\max(\omega(\mathcal{G}),\beta(\mathcal{G})) = \infty$. Since $\mathcal G$ is monotone, if $\omega(\mathcal{G})=\infty$ then $\mathcal G$ contains (the $0$-subdivision of) every clique, and if $\beta(\mathcal{G})=\infty$ then $\mathcal{G}$ contains all bipartite graphs and thus the $1$-subdivision of every clique. In any case Lemma~\ref{lem:nd_subdivisions} implies that $\mathcal{G}$ is somewhere dense, contradicting the assumptions.
\end{proof}

Theorem~\ref{thm:subs_monotone_intro} follows immediately.
\begin{corollary}[Theorem~\ref{thm:subs_monotone_intro}, restated]\label{cor:subs_monotone_intro}
Let $\mathcal{H}$ and $\mathcal{G}$ be monotone graph classes and assume that ETH holds. Then $\#\subsprob(\mathcal{H} \to \mathcal{G})$ is fixed-parameter tractable if $\m(\mathcal{H})<\infty$ or $\mathcal{G}$ is nowhere dense; otherwise $\#\subsprob(\mathcal{H} \to \mathcal{G})$ is $\#\W[1]$-complete and cannot be solved in time $f(|H|)\cdot |G|^{o(|V(H)|/ \log(|V(H)|))}$ for any function $f$.
\end{corollary}
\begin{proof}
If $\mathcal{H}$ is monotone then $\mathcal{H}$ is hereditary and Theorem~\ref{thm:subs_hereditary_intro} applies. The union of the first row and the first column of Table~\ref{tab:resultsSubsHereditary} yield the tractable case; the union of the remaining entries yield the intractable case and the lower bounds.
\end{proof}

We conclude this section with a remark.
\begin{remark}\label{rem:realFPT}
Let $\mathcal{H}$ and $\mathcal{G}$ be the classes of graphs of degrees bounded by $2$ and $3$, respectively. Then $\#\subsprob(\mathcal{H}\to\mathcal{G})$ subsumes the $\#\mathrm{P}$-hard problem of counting Hamiltonian cycles in $3$-regular graphs. Since both classes are monotone (and thus also hereditary), since $\mind(\mathcal{H})=\infty$, and since classes of bounded degree graphs are nowhere dense (see e.g.\ \cite{GroheKS17}), this shows that the FPT entry in Table~\ref{tab:resultsSubsHereditary} cannot be strengthened to P unless $\#\mathrm{P}=\mathrm{P}$.
\end{remark}

\section{Counting Induced Subgraphs}
\label{sec:inds}
This section is devoted to the proofs of Theorem~\ref{thm:intro_is} and Theorem~\ref{thm:indsubs_hereditary_intro}. We begin in Section~\ref{sec:inds_1} by analysing the problem of counting independent sets and proving Theorem~\ref{thm:intro_is}; this is the most technical part. We then prove Theorem~\ref{thm:indsubs_hereditary_intro} in Section~\ref{sec:inds_2}.

\subsection{Counting Independent Sets: Proof of Theorem~\ref{thm:intro_is}}\label{sec:inds_1}
Given a class of graphs $\mathcal{G}$, the problem $\#\is(\mathcal{G})$ asks, on input $k \in \N$ and a graph $G\in \mathcal{G}$, to compute the number of independent sets of size $k$ (also called $k$-\emph{independent sets}) in $G$. In this section we prove hardness results for $\#\is(\mathcal{G})$ and leverage them to $\#\indsubsprob(\mathcal{H} \to \mathcal{G})$. To this end we will rely on subgraphs induced by sets of edges; they play a role similar to that of fractured graphs in Section~\ref{sec:subgraphs}.
Given a graph $F$ and a set $A \subseteq E(F)$, we denote the subgraph $(V(F),A)$ by $F[A]$. For what follows observe that, for any $A\subseteq E(F)$, the identity function on $V(F)$, which we denote by~$\mathsf{id}_F$, is a surjective $F$-colouring of $F[A]$.
Now recall Definition~\ref{def:tensor}.
We start with the following simple variation of Lemma~\ref{lem:main_tensor_subs}.
\begin{lemma}\label{fact:main_tensor_es}
Let $r \in \No$, let $H$ be a graph without isolated vertices, let $G$ be an $H$-coloured graph, and let $A \subseteq E(H^r)$. Then $(G^r,c^r) \times (H^r[A],\mathsf{id}_{H^r})$
is a subgraph of $K_{|V(G)|}^r$.
\end{lemma}
\begin{proof}
Let $n=|V(G)|$. First, note that $(G^r,c^r) \times (H^r,\mathsf{id}_{H^r})=(G^r,c^r)$, and by construction $(G^r,c^r)$ is a subgraph of $K_n^r$. Next, for every $A\subseteq E(H^r)$ the graph $(G^r,c^r) \times (H^r[A],\mathsf{id}_{H^r})$ is obtained from $(G^r,c^r)$ by deleting edges --- specifically, for every $e=\{u,v\} \in E(H^r)\setminus A$, delete from $G^r$ all edges between vertices coloured with $u$ and vertices coloured with $v$. Thus $(G^r,c^r) \times (H^r[A],\mathsf{id}_{H^r})$ is a subgraph of $K_n^r$ too.
\end{proof}
Recall that $\#\colmatch(\mathcal{G})$, the problem of counting edge-colourful $k$-matchings, was the key subproblem in the hardness proofs for $\#\subsprob(\scH \to \scG)$ --- see Section~\ref{sec:subs_2}. In the case of $\#\indsubsprob(\scH \to \scG)$, the key subproblem turns out to be that of counting vertex-colourful independent sets.
Let $G$ be a graph and let $c:V(G)\to \{1,\ldots,k\}$ be a coloring of $V(G)$. A set $U \subseteq V(G)$ is \emph{vertex-colorful} if for every colour in $\{1,\ldots,k\}$ there is precisely one element of $U$ with that colour.
\begin{definition}[$\#\colis(\mathcal{G})$]
Let $\mathcal{G}$ be a class of graphs. The problem $\#\colis(\mathcal{G})$ asks, on input $k \in \N$, a graph $G\in\mathcal{G}$, and a $k$-coloring $c$ of $V(G)$, to compute the number of vertex-colorful $k$-independent sets in $G$. The problem is parameterised by $k$.
\end{definition}
Our goal is to show that $\#\colis(\mathcal{G})$ is intractable whenever $\mathcal{G}$ is monotone and somewhere dense. As for $\#\colmatch(\mathcal{G})$ in Section~\ref{sec:subs_2}, the reduction relies on solving a system of linear equations. Let $H$ be a graph. The square matrix $N_H$ has its rows and columns indexed by the subsets of $E[H]$, and its entries satisfy
\begin{equation}\label{def:N_H}
    N_H[A,B] = \#\homs{(H[A],\id_H)}{(H[B],\id_H)}\,.
\end{equation}
Similarly to the matrix $M_H$ in Section~\ref{sec:subs_2}, the following was established in prior work:
\begin{lemma}[\cite{DorflerRSW22}]\label{lem:N_H_triangular}
For each graph $H$, the matrix $N_H$ is nonsingular.
\end{lemma}

We are now able to establish intractability of $\#\colis(\mathcal{G})$.

\begin{lemma}\label{lem:colis_hard}
Let $\mathcal{G}$ be a monotone somewhere dense class of graphs. Then 
$\#\colis(\mathcal{G})$ is $\#\mathsf{W}[1]$-complete and, assuming ETH, cannot be solved in time $f(k)\cdot |G|^{o(k/\log k)}$ for any function $f$.
\end{lemma}
\begin{proof}
The proof is similar to that of Lemma~\ref{lem:main_colmatch}. First, since $\mathcal{G}$ is monotone and somewhere dense, by Lemma~\ref{lem:nd_subdivisions} there exists $r\in\No$ such that $G^r \in \mathcal{G}$ for every $G \in \all$. Second, let $\mathcal{H}$ be a class of $3$-regular expander graphs. By theorems~\ref{thm:cphomshard} and~\ref{thm:cybt}, $\#\cphomsprob(\mathcal{H} \to \all)$ is $\#\W[1]$-hard and assuming ETH cannot be solved in time $f(|H|)\cdot |G|^{o(|H|/\log |H|)}$ for any function $f$.
We show a parameterized reduction from $\#\cphomsprob(\mathcal{H} \to \all)$ to $\#\colis(\scG)$.

Let $(H,(G,c))$ be the input to $\#\cphomsprob(\mathcal{H} \to \all)$. Our reduction starts by constructing $H^r$ and $(G^r,c^r)$, which by Lemma~\ref{lem:hom_subdiv} satisfy
\[\#\homs{(H,\id_H)}{(G,c)}=\#\homs{(H^r,\id_{H^r})}{(G^r,c^r)}\,.\]
Let $k=|V(H^r)|$; clearly $k\in O(|H|)$ since $r$ is a constant independent of $H$. Our goal is to use the oracle for $\#\colis(\mathcal{G})$ to compute $\#\homs{(H^r,\id_{H^r})}{(G^r,c^r)}$. From now on we view surjectively $H^r$-coloured graphs $(\Tilde{G},\Tilde{c})$ also as vertex-coloured graphs with colouring $\Tilde{c}$. This allows us to invoke~\cite[Lemma 8]{DorflerRSW22} and obtain what follows.\footnote{In~\cite{DorflerRSW22} the number of colourful $k$-independent sets in a surjectively $H^r$-coloured graph $\hat{G}$ is denoted by $\#\mathsf{cp}\text{-}\mathsf{IndSub}(\Phi \to_{H^r} \hat{G})$, where $\Phi$ is the graph property of being an independent set.}

First, there is a unique function $\hat{a}$ from subsets of $E[H^r]$ to rationals such that, for every surjectively $H^r$-coloured graph $(\Tilde{G},\Tilde{c})$, the number of vertex-colourful $k$-independent sets in $(\Tilde{G},\Tilde{c})$ equals
\begin{equation}\label{eq:colIS1}
    \sum_A \hat{a}(A) \cdot \#\homs{(H^r[A],\mathsf{id}_{H^r})}{(\Tilde{G},\Tilde{c})}\,,
\end{equation}
where the sum is over all subsets of $E[H^r]$. Additionally,
\begin{equation}\label{eq:colIS2}
    \hat{a}(E(H^r)) = \pm \hat{\chi} \neq 0\,, 
\end{equation}
where $\hat{\chi}$ is the so-called alternating enumerator for the graph property of being an independent set --- we omit the definition since the only property needed for $\hat{\chi}$ is it being easily computable and non-zero (see~\cite{DorflerRSW22}).

Now consider~\eqref{eq:colIS1} with $(\tilde G,\tilde c)=(G^r,c^r) \times (H^r[B],\mathsf{id}_{H^r})$ and apply Lemma~\ref{lem:tensor_linear_subgraphs}. We deduce that the number of vertex-colourful $k$-independent sets in $(\Tilde{G},\Tilde{c})$ is
\begin{align*}
\sum_A \hat{a}(A)\cdot \#\homs{(H^r[A],\mathsf{id}_{H^r})}{(G^r,c^r)} \cdot \#\homs{(H^r[A],\mathsf{id}_{H^r})}{(H^r[B],\mathsf{id}_{H^r})}. 
\end{align*}
By Lemma~\ref{fact:main_tensor_es}, for every $B\subseteq E(H^r)$ of $H^r$ the graph $(G^r,c^r) \times (H^r[B],\mathsf{id}_{H^r})$ is a subgraph of the $r$-subdivision of a complete graph; by the monotonicity of $\mathcal{G}$ and by the choice of $r$ this implies $(G^r,c^r) \times (H^r[B],\mathsf{id}_{H^r}) \in \scG$, see Lemma~\ref{lem:nd_subdivisions}. 
Thus, as in the proof of Lemma~\ref{lem:main_colmatch}, by using an oracle for $\#\colis(\scG)$ we can construct in FPT time a system of linear equations whose matrix $N_{H^r}$ is nonsingular by Lemma~\ref{lem:N_H_triangular}. Since $\hat{a}(E(H^r))\neq 0$ by~(\ref{eq:colIS2}), solving this system enables us to compute 
\[\#\homs{(H^r[E(H^r)],\mathsf{id}_{H^r})}{(G^r,c^r)} =\#\homs{(H^r,\mathsf{id}_{H^r})}{(G^r,c^r)} \,,\]
concluding the proof.
\end{proof}

With the above hardness results for $\#\colis(\mathcal{G})$, we can finally prove complexity dichotomies for its non-coloured counterpart $\#\is(\mathcal{G})$. We start by porting Lemma~\ref{lem:colis_hard} from $\#\colis(\mathcal{G})$ to $\#\is(\mathcal{G})$.

\begin{theorem}\label{thm:main_is}
Let $\mathcal{G}$ be a monotone somewhere dense class of graphs. Then $\#\is(\mathcal{G})$ is $\#\mathsf{W}[1]$-hard and, assuming ETH, cannot be solved in time $f(k)\cdot |G|^{o(k/\log k)}$ for any function $f$.
\end{theorem}
\begin{proof}
Almost identical to the proof of Theorem~\ref{thm:main_matchings}: when $\mathcal G$ is monotone, $\#\colis(\mathcal{G})$ can be reduced in FPT time to $\#\is(\mathcal{G})$ via inclusion-exclusion while preserving the parameter (see, for instance, \cite[Lemma~1.34]{Curticapean15}), and the claim then follows by Lemma~\ref{lem:colis_hard}.
\end{proof}

\noindent We can finally prove Theorem~\ref{thm:intro_is} as a simple corollary.
\begin{corollary}[Theorem~\ref{thm:intro_is}, restated]\label{xxxIS}
Let $\mathcal{G}$ be a monotone class of graphs and assume that ETH holds. Then $\#\is(\mathcal{G})$ is fixed-parameter tractable if and only if $\mathcal{G}$ is nowhere dense. In particular, if $\mathcal{G}$ is nowhere dense then $\#\is(\mathcal{G})$ can be solved in time $f(k)\cdot|V(G)|^{1+o(1)}$ for some computable function~$f$; otherwise $\#\is(\mathcal{G})$ cannot be solved in time $f(k)\cdot |G|^{o(k/\log k)}$ for any  function $f$. 
\end{corollary}
\begin{proof}
Immediate by Theorem~\ref{thm:sparsity} and Theorem~\ref{thm:main_is}.
\end{proof}

We conclude with a remark.
\begin{remark}\label{rem:realFPTind} Corollary~\ref{xxxIS} cannot be strengthened to polynomial-time tractability of $\#\is(\mathcal{G})$ when $\mathcal{G}$ is nowhere dense and monotone, unless $\#\mathrm{P}=\mathrm{P}$: graphs of degree at most $3$ form such a class, yet counting independent sets in them is $\#\mathrm{P}$-hard~\cite{Greenhill00}.
\end{remark}

\subsection{Counting Induced Subgraphs: Proof of Theorem~\ref{thm:indsubs_hereditary_intro}}\label{sec:inds_2}
Equipped with our complexity dichotomy for $\#\is(\mathcal{G})$, we can now prove our complexity dichotomies for $\#\indsubsprob(\mathcal{H} \to \mathcal{G})$. First, we consider the case that $\mathcal H$ is monotone.
\begin{corollary}\label{cor:indsubs_mono}
Let $\mathcal{H}$ and $\mathcal{G}$ be monotone graph classes and assume that ETH holds. Then $\#\indsubsprob(\mathcal{H} \to \mathcal{G})$ is fixed-parameter tractable if $\mathcal{H}$ is finite or $\mathcal{G}$ is nowhere dense; otherwise $\#\indsubsprob(\mathcal{H} \to \mathcal{G})$ is $\#\W[1]$-complete and cannot be solved in time $f(|H|)\cdot |G|^{o(|V(H)|/ \log(|V(H)|))}$ for any function $f$.
\end{corollary}
\begin{proof}
If $\mathcal{H}$ is finite then  $\#\indsubsprob(\mathcal{H} \to \mathcal{G})$ is clearly in polynomial time (and thus fixed-parameter tractable) since the brute-force algorithm runs in time $O(|G|^{|H|})$. If $\mathcal{G}$ is nowhere dense then the fixed-parameter tractability follows by Theorem~\ref{thm:sparsity}. Finally, if $\mathcal{H}$ is monotone and infinite then it contains all independent sets, and thus $\#\indsubsprob(\mathcal{H} \to \mathcal{G})$ subsumes $\#\is(\mathcal{G})$; in which case Theorem~\ref{thm:main_is} yields the lower bound for somewhere dense $\mathcal{G}$.
\end{proof}

Next, we consider the case that $\mathcal H$ is hereditary. We obtain a refined complexity classification that subsumes the one of Corollary~\ref{cor:indsubs_mono} and yields Theorem~\ref{thm:indsubs_hereditary_intro}. 
\begin{table}[t]
   \begin{tabularx}{\textwidth}{l|ccc}
          \begin{tabular}{c}
             $~$
        \end{tabular}&\begin{tabular}{c}
             $\mathcal{G}$
             nowhere dense\\
             $~$
        \end{tabular} &\begin{tabular}{c}
             $\mathcal{G}$
             somewhere dense\\
             $\omega(\mathcal{G})=\infty$
        \end{tabular}& \begin{tabular}{c}
             $\mathcal{G}$
             somewhere dense\\
             $\omega(\mathcal{G})< \infty$, 
             ${\alpha}(\mathcal{G})=\infty$
        \end{tabular}\\
        & \\[\dimexpr-\normalbaselineskip+5pt]
        \hline
        & \\[\dimexpr-\normalbaselineskip+5pt]
        \begin{tabular}{l}
             $\mathcal{H}$ finite
        \end{tabular} & P & P & P  \\
        & \\[\dimexpr-\normalbaselineskip+5pt]
        \hline
        & \\[\dimexpr-\normalbaselineskip+5pt]
        \begin{tabular}{l}
             $\alpha(\mathcal{H})= \infty$
        \end{tabular} & FPT & \begin{tabular}{c}
             $\#\W[1]$-hard\\ not in $f(k)\cdot n^{o(k)}$
        \end{tabular} & \begin{tabular}{c}
             $\#\W[1]$-hard\\ not in $f(k)\cdot n^{o(k/\log k)}$
        \end{tabular}  \\
        & \\[\dimexpr-\normalbaselineskip+5pt]
        \hline
        & \\[\dimexpr-\normalbaselineskip+5pt]
        \begin{tabular}{l}
            $\alpha(\mathcal{H})< \infty$\\
            $\omega(\mathcal{H})= \infty$
        \end{tabular} & P & \begin{tabular}{c}
             $\#\W[1]$-hard\\ not in $f(k)\cdot n^{o(k)}$
        \end{tabular} & P  
        \end{tabularx}
     \caption[caption]{\small The complexity of $\#\indsubsprob(\mathcal{H}\to \mathcal{G})$ for hereditary $\mathcal{H}$ and monotone $\mathcal{G}$. P and FPT stand respectively for polynomial-time tractability and fixed-parameter tractability, and hard means $\#\W[1]$-hard and without an algorithm running in time $f(k)\cdot n^{o(k/ \log(k))}$ for any function $f$ unless ETH fails, where $k=|V(H)|$ and $n=|V(G)|$. The FPT entry cannot be strengthened to P unless $\mathrm{P}=\#\mathrm{P}$, see Remark~\ref{rem:realFPTind}.
}
\label{tab:resultsIndSubsHereditary}
\end{table}

\begin{theorem}[Theorem~\ref{thm:indsubs_hereditary_intro}, restated]
Let $\mathcal{H}$ and $\mathcal{G}$ be graph classes such that $\mathcal{H}$ is hereditary and $\mathcal{G}$ is monotone. Then Table~\ref{tab:resultsIndSubsHereditary} exhaustively classifies the complexity of $\#\indsubsprob(\mathcal{H} \to \mathcal{G})$.
\end{theorem}
\begin{proof}
The cases for $\mathcal{G}$ and $\mathcal{H}$ in Table~\ref{tab:resultsIndSubsHereditary} are mutually exclusive and exhaustive by Ramsey's Theorem (Theorem~\ref{thm:ramsey}). Let us then prove the entries of Table~\ref{tab:resultsIndSubsHereditary}.

The first row holds since for finite $\mathcal{H}$ the brute-force algorithm runs in polynomial time, and the FPT result follows from Theorem~\ref{thm:sparsity}.
For the intractability results in the second column, note that since $\mathcal{G}$ is monotone and infinite then $\mathcal{G}=\all$, and since $\mathcal{H}$ is hereditary, the cases $\alpha(\mathcal{H})=\infty$ and $\omega(\mathcal{H})=\infty$ subsume respectively $\#\is(\all)$ and $\#\clique(\all)$. Both are canonical $\#\W[1]$-hard problems and cannot be solved in time $f(k)\cdot n^{o(k)}$ unless ETH fails~\cite{Chenetal05,Chenetal06}.\footnote{The lower bound in~\cite{Chenetal05,Chenetal06} applies to counting $k$-cliques, and we note that counting $k$-cliques and counting $k$-independent sets are interreducible by taking the complement of the host.}
The intractability results in the third column follows from Theorem~\ref{thm:main_is} since $\mathcal{H}$ being hereditary and $\alpha(\mathcal{H})=\infty$ imply that $\#\indsubsprob(\mathcal{H} \to \mathcal{G})$ subsumes $\#\is(\mathcal{G})$. 

It remains to prove the first and the third entry of the third row. Note that both entries assume $\omega(\mathcal{G}) < \infty$ and $\alpha(\mathcal{H}) < \infty$. Let then $(H,G)$ be the input. If $\omega(H)>\omega(\scG)$ then we can immediately return~$0$. Otherwise $|V(H)|\le R(\omega(\scG),\alpha(\scH)) < \infty$, where $R$ is Ramsey's function (see Theorem~\ref{thm:ramsey}), and the brute-force algorithm runs in polynomial time. 
\end{proof}

\section{Outlook}
Due to the absence of a general dichotomy~\cite{RothW20}, the following two directions are evident candidates for future analysis.

\paragraph*{Hereditary Host Graphs.}
Is there a way to refine our classifications to hereditary $\mathcal{G}$? 
While such results would naturally be much stronger, we point out that a classification of general first-order (FO) model-checking and model-counting in hereditary graphs is wide open. Concretely, even if $\mathcal{H}=\all$, it currently seems elusive to obtain criteria for hereditary $\mathcal{G}$ which, if satisfied, yield fixed-parameter tractability of  $\#\subsprob(\mathcal{H}\to\mathcal{G})$,  $\#\indsubsprob(\mathcal{H}\to\mathcal{G})$, and $\#\homsprob(\mathcal{H}\to\mathcal{G})$ and which, if not satisfied, yield $\#\W[1]$-hardness of those problems. In a nutshell, the problem is that there are arbitrarily dense hereditary classes of host graphs for which those problems, and even the much more general FO-model counting problem, become tractable; a trivial example is given by $\mathcal{G}$ being the class of all complete graphs. See~\cite{GajarskyHLOORS15,Ganianetal15,GenietT22} for recent work on specific hereditary hosts and~\cite{GajarskyHOLR20,BonnetKTW22} for general approaches to understand FO model checking on dense graphs.

\paragraph*{Arbitrary Pattern Graphs.}
Can we refine our classifications to arbitrary classes of patterns $\mathcal{H}$, given that we stay in the realm of monotone classes of hosts $\mathcal{G}$? We believe this question is the most promising direction for future research. While a sufficient and necessary criterion for the fixed-parameter tractability of, say $\#\subsprob(\mathcal{H} \to \mathcal{G})$, must depend on the set of forbidden subgraphs of $\mathcal{G}$, we conjecture that the structure of monotone somewhere dense graph classes is rich enough to allow for an explicit combinatorial description of such a criterion. In fact, such criteria have already been established for some specific classes of host graphs, e.g.\ bipartite graphs~\cite{CurticapeanM14} and degenerate graphs~\cite{BressanR21}.

\newpage

\bibliographystyle{plainurl}
\bibliography{patterncountbib}

\end{document}